\documentclass[12pt,prd]{revtex4-1}
\usepackage{CJK}
\usepackage[T1]{fontenc}
\usepackage[left=1.5cm,top=2cm,right=1.5cm,bottom=4cm]{geometry} %Márgenes del documento

\usepackage[utf8]{inputenc} %Escribir símbolos no anglosajones como á, ü, ê, ò, etc.

\usepackage{array}
\usepackage{textcomp}
\usepackage{multirow}
\usepackage{amsmath}
\usepackage{graphicx}
\usepackage{hhline}
\usepackage{breqn}
\usepackage{comment}
\usepackage{adjustbox}
\usepackage{support-caption}
\usepackage{subcaption}
\usepackage{appendix}

\usepackage{amsmath}
\usepackage{amsfonts}
\usepackage{amssymb}
\usepackage{cancel} %Barra de cancelación
\usepackage{youngtab}
\usepackage{xfrac}

\usepackage{pstricks}
\usepackage{color}
\usepackage{graphicx}
\usepackage{feynmp} %Feynman diagrams package
\usepackage{float} %Opción de [H] para imágenes. 

\newcommand{\be}{\begin{equation}}
\newcommand{\ee}{\end{equation}}
\newcommand{\bea}{\begin{eqnarray}}
\newcommand{\eea}{\end{eqnarray}}

\usepackage{booktabs}% http://ctan.org/pkg/booktabs

\begin{document}

\title{Explaining muon $g-2$ anomaly in a non-universal $U(1)_{X}$ extended SUSY theory}
\author{J.S. Alvarado} \thanks{jsalvaradog@unal.edu.co}
\author{M.A. Bulla} \thanks{mabullar@unal.edu.co}
\author{D.G. Martinez} \thanks{dgmartinezg@unal.edu.co }
\author{R. Martinez. } \thanks{remartinezm@unal.edu.co}

\affiliation{Departamento de Física$,$ Universidad Nacional de Colombia\\
 Ciudad Universitaria$,$ K. 45 No. 26-85$,$ Bogot\'a D.C.$,$ Colombia}
\date{\today}

\begin{abstract}
    A non-universal $U(1)_{X}$ extension to the Standard Model composed of two scalar doublets and two scalar singlets together with three additional quark singlets and two lepton singlets and three generations of right-handed and Majorana neutrinos is made to explain lepton mass hierarchy, neutrino masses via inverse seesaw mechanism and muon anomalous magnetic moment in an anomaly free framework. In the present model, exotic and Standard Model particles acquire mass thanks to vacuum expectation values at different scales, yet the electron and the lightest neutrino are tree level massless but massive at one-loop level. By considering a numerical exploration and under the constraint of the Higgs mass, neutrino mass differences and PMNS matrix, it is found that only contributions due to exotic neutrinos interacting with charged scalars are relevant to muon $g-2$, though they are negative. Thus, the SUSY extension is considered and it is found that  muon $g-2$ can be explained by allowing $U(1)_{X}$ vacuum expectation values to lie in the TeV scale thanks to SUSY soft-breaking interactions for at least $\sim 10^{5}$ GeV masses. Thus, the contribution due to exotic neutrinos interacting with $W$ gauge bosons is positive and no longer negligible which added to all other contributions might explain the anomaly. 
    
\end{abstract}

\maketitle

\section{Introduction}

One of the most important results in QED comes from the measurement of the electron magnetic moment which leads to an agreement of up to thirteen significant figures to the fine structure constant \cite{electrong-2} when including hadronic and electroweak corrections correctly predicted by the Standard Model (SM) \cite{weak}\cite{salam}\cite{weinberg}. However, muon anomalous magnetic moment does not agree with experiments, presenting a current deviation of $ \Delta a_{\mu}= (27.9 \pm 7.6)\times 10^{-10} $ at $3.7\sigma$ \cite{DeltaAmeasured} with an elevated theoretical uncertainty which is believed to come from the inputs of the hadronic loop contributions \cite{hadronic1} since the other two have been confirmed with high precision already \cite{electroweak1}\cite{electroweak2}. Nevertheless, reducing the uncertainty gap is one of the most important expectations in the near future, for example Fermilab E989 \cite{experiment1} expects to measure nearly a $5\sigma$ deviation and similarly by  J-PARC \cite{experiment2}. This $a_{\mu}=\frac{g_{\mu}-2}{2}$ has become one of the mysteries to explain through models beyond the standard model which mostly includes new exotic particles, flavor changing processes as part of different higher symmetry groups such as 331 models \cite{model1}, $L_{\mu}-L_{\tau}$ $U(1)$ symmetries, $U(1)$ extensions \cite{model5} and anomaly mediation \cite{model4} among other new physics scenarios \cite{model6}    . \\

Inside the loop might be particles with a vector(scalar) and axial(pseudoscalar) for a gauge(scalar) boson connecting the external lines which implies a positive (negative) contribution. In general, the theoretical contributions have been already calculated in \cite{generalintegrals1}\cite{generalintegrals2}\cite{generalintegrals3} which corresponds to a general interaction lagrangian $ \mathcal{L}_{int}= g_{s}\bar{\psi}\mu \phi+ig_{p}\bar{\psi}\gamma^{5}\mu \phi $ and $ \mathcal{L}_{int}= g_{v}\bar{\psi}\gamma
^{\mu}\mu Z'_{\mu}+ig_{p}\bar{\psi}\gamma^{5}\gamma^{\mu}\mu Z'_{\mu} $ for scalar and gauge bosons contributions although the particular interaction of muons with the Higgs boson is of special relevance despite its highly suppressed value because recent measurements still leave a gap for possible new physics contributions \cite{htomumu} whose latest report by ATLAS experiment is a signal strength of $\mu=1.2 \pm 0.6$ corresponding to a $2\sigma$ significance in relation to the no-signal hypothesis. \\

The present work, presents an anomaly free SM $U(1)_{X}$ extension containing two scalar doublets, two scalar singlets, tree singlet quarks, two singlet leptons and tree generations of right-handed and Majorana neutrinos in section \ref{themodel} where gauge boson, scalar and lepton mass eigenvalues and the corresponding rotation matrices are obtained. Next, in section \ref{nosusycontributions} the interactions and contributions to muon $g-2$ are shown with a sample parameter choice of values that allows to see the model general behavior. Since the net contribution turns out to be negative, we explore its supersymmetric generalization in section \ref{SUSYgen} where only scalar mass eigenstates are revisited. Their contributions are calculated in section \ref{SUSYcontributions} where we show that by adding all contributions the experimental muon $g-2$ can be reproduced. Finally, it is presented a small section regarding conclusions in \ref{conclu}.

\section{the $U(1)_{X}$ extension} \label{themodel}
By considering an appropriate choice of $X$ charges and a $\mathbb{Z}_{2}$ symmetry to better restrict mass matrix textures we consider a model with two scalar doublets and two scalar singlets where one of them has a null vacuum expectation value (VEV). Likewise, the fermion content has a non-universal $X$-charge assignation and additional three quark singlets and two lepton singlets as well as three generations of right-handed and Majorana neutrinos to explain neutrino masses via inverse seesaw mechanism as will be seen later. The  particle content and their quantum numbers are specified in tables \ref{scalarlist} and \ref{fermionlist}.

\begin{table}[H]
    \centering
    \begin{tabular}{|ccc|ccc|} \hline \hline
        Scalar Doublets & & & Scalar Singlets & &  \\
         & $X^{\pm}$ & $Y$ & & $X^{\pm}$ & $Y$   \\ \hline \hline
    $\small{\phi_{1}=\begin{pmatrix}\phi_{1}^{+}\\\frac{h_{1}+v_{1}+i\eta_{1}}{\sqrt{2}}\end{pmatrix}}$ & $\sfrac{+2}{3}^{+}$ & $+1$ & $\chi=\frac{\xi_{\chi}+v_{\chi}+i\zeta_{\chi}}{\sqrt{2}}$	&	$\sfrac{-1}{3}^{+}$	&	$0$	\\
    $\small{\phi_{2}=\begin{pmatrix}\phi_{2}^{+}\\\frac{h_{2}+v_{2}+i\eta_{2}}{\sqrt{2}}\end{pmatrix}}$&$\sfrac{+1}{3}^{-}$&$+1$& $\sigma=\frac{\sigma+i\zeta_{\sigma}}{\sqrt{2}} $ &$ \sfrac{-1}{3}^{-} $ & $ 0 $ \\\hline
    \end{tabular}
\caption{Model scalar particle content, $X$-charge, $\mathbb{Z}_{2}$ parity and hypercharge}
\label{scalarlist}
\end{table}

\begin{table}[H]
\centering
\begin{tabular}{|cccc|ccc|}
\hline\hline
Quarks	&	$X$	&$\mathbb{Z}_{2}$&&	Leptons	&	$X$&$\mathbb{Z}_{2}$	\\ \hline 
$q^{1}_{L}=\left(\begin{array}{c}U^{1} \\ D^{1} \end{array}\right)_{L}$
	&	$+1/3$	&$+$	&&
$\ell^{e}_{L}=\left(\begin{array}{c}\nu^{e} \\ e^{e} \end{array}\right)_{L}$
	&	$0$	&$+$	\\
$q^{2}_{L}=\left(\begin{array}{c}U^{2} \\ D^{2} \end{array}\right)_{L}$
	&	$0$	&$-$	&&
$\ell^{\mu}_{L}=\left(\begin{array}{c}\nu^{\mu} \\ e^{\mu} \end{array}\right)_{L}$
	&	$0$	&$+$		\\
$q^{3}_{L}=\left(\begin{array}{c}U^{3} \\ D^{3} \end{array}\right)_{L}$
	&	$0$	&$+$	&&
$\ell^{\tau}_{L}=\left(\begin{array}{c}\nu^{\tau} \\ e^{\tau} \end{array}\right)_{L}$
	&	$-1$	&$+$	\\   \hline\hline

\begin{tabular}{c}$U_{R}^{1,3}$\\$U_{R}^{2}$\\$D_{R}^{1,2,3}$\end{tabular}	&	 
\begin{tabular}{c}$+2/3$\\$+2/3$\\$-1/3$\end{tabular}	&
\begin{tabular}{c}$+$\\$-$\\$-$\end{tabular}	&&
\begin{tabular}{c}$e_{R}^{e,\tau}$\\$e_{R}^{\mu}$\end{tabular}	&	
\begin{tabular}{c}$-4/3$\\$-1/3$\end{tabular}	&	
\begin{tabular}{c}$-$\\$-$\end{tabular}\\   \hline \hline 

\multicolumn{3}{|c}{Non-SM Quarks}	&&	\multicolumn{3}{c|}{Non-SM Leptons}	\\ \hline \hline
\begin{tabular}{c}$T_{L}$\\$T_{R}$\end{tabular}	&
\begin{tabular}{c}$+1/3$\\$+2/3$\end{tabular}	&
\begin{tabular}{c}$-$\\$-$\end{tabular}	&&
\begin{tabular}{c}$\nu_{R}^{e,\mu,\tau}$\\$N_{R}^{e,\mu,\tau}$\end{tabular} 	&	
\begin{tabular}{c}$1/3$\\$0$\end{tabular}	&	
\begin{tabular}{c}$-$\\$-$\end{tabular}\\
$J^{1,2}_{L}$	&	  $0$ 	&$+$	&&	$E_{L},\mathcal{E}_{R}$	&	$-1$	&$+$	\\
$J^{1,2}_{R}$	&	 $-1/3$	&$+$	&&	$\mathcal{E}_{L},E_{R}$	&	$-2/3$	&$+$	\\ \hline \hline
\end{tabular}
\caption{Model fermion particle content, $X$-charge, $\mathbb{Z}_{2}$ parity and hypercharge.}
\label{fermionlist}
\end{table}

When additional symmetries are included in a theory, there is always the risk of inducing chiral anomalies so the $X$-charges were chosen to satisfy the anomaly cancellation equations shown in (\ref{eq:Chiral-anomalies}) so the model can be anomaly-free and renormalizability can be ensured. 

\begin{widetext}
\begin{eqnarray}
\label{eq:Chiral-anomalies}
\left[\mathrm{\mathrm{SU}(3)}_{C} \right]^{2} \mathrm{\mathrm{U}(1)}_{X} \rightarrow & A_{C} &= \sum_{Q}X_{Q_{L}} - \sum_{Q}X_{Q_{R}}	\nonumber	\\
\left[\mathrm{\mathrm{SU}(2)}_{L} \right]^{2} \mathrm{\mathrm{U}(1)}_{X} \rightarrow & A_{L}  &= \sum_{\ell}X_{\ell_{L}} + 3\sum_{Q}X_{Q_{L}}	\nonumber	\\
\left[\mathrm{\mathrm{U}(1)}_{Y} \right]^{2}   \mathrm{\mathrm{U}(1)}_{X} \rightarrow & A_{Y^{2}}&=
	\sum_{\ell, Q}\left[Y_{\ell_{L}}^{2}X_{\ell_{L}}+3Y_{Q_{L}}^{2}X_{Q_{L}} \right]\nonumber	%\\ &&
	- \sum_{\ell,Q}\left[Y_{\ell_{R}}^{2}X_{L_{R}}+3Y_{Q_{R}}^{2}X_{Q_{R}} \right]	\nonumber	\\
\mathrm{\mathrm{U}(1)}_{Y}   \left[\mathrm{\mathrm{U}(1)}_{X} \right]^{2} \rightarrow & A_{Y}&=
	\sum_{\ell, Q}\left[Y_{\ell_{L}}X_{\ell_{L}}^{2}+3Y_{Q_{L}}X_{Q_{L}}^{2} \right]\nonumber	%\\ &&
	- \sum_{\ell, Q}\left[Y_{\ell_{R}}X_{\ell_{R}}^{2}+3Y_{Q_{R}}X_{Q_{R}}^{2} \right]	\nonumber	\\
\left[\mathrm{\mathrm{U}(1)}_{X} \right]^{3} \rightarrow & A_{X}&=
	\sum_{\ell, Q}\left[X_{\ell_{L}}^{3}+3X_{Q_{L}}^{3} \right]	%\\ &&
	- \sum_{\ell, Q}\left[X_{\ell_{R}}^{3}+3X_{Q_{R}}^{3} \right] 	\nonumber	\\	
\left[\mathrm{Grav} \right]^{2}   \mathrm{\mathrm{U}(1)}_{X} \rightarrow & A_{\mathrm{G}}&=
	\sum_{\ell, Q}\left[X_{\ell_{L}}+3X_{Q_{L}} \right]%\\ &&
	- \sum_{\ell, Q}\left[X_{\ell_{R}}+3X_{Q_{R}} \right],
\end{eqnarray}
\end{widetext}
\noindent
where subscripts $Q$ and $\ell$ represent quarks and leptons respectively and $L$ and $R$ represent left and right chirality respectively. Despite Majorana particles do not contribute to anomaly equations they were included to implement an inverse seesaw mechanism (ISS) for neutrino mass generation \cite{inverseseesaw}. Additional scalar singlets were included to explain exotic fermion masses via $\chi$ singlet interactions and light fermion masses via $\sigma$ interactions at one-loop level. Besides, the $\chi$ singlet breaks the $U(1)_{X}$ symmetry when a TeV VEV is acquired to recover the SM gauge symmetry via the following spontaneous symmetry breaking chain:
\begin{eqnarray}
    \mathrm{SU(3)}_{C}\otimes
    \mathrm{SU(2)}_{L}\otimes 
    \mathrm{U(1)}_{Y} \otimes 
    \mathrm{U(1)}_{X} &\overset{\chi}{\longrightarrow}&\\
    \mathrm{SU(3)}_{C}\otimes
    \mathrm{SU(2)}_{L}\otimes 
    \mathrm{U(1)}_{Y} &\overset{\Phi}{\longrightarrow} &
    \mathrm{SU(3)}_{C}\otimes
    \mathrm{U(1)}_{Q}. \nonumber
\end{eqnarray}

\subsection{Gauge Boson Masses}
Due to the new symmetry, a new gauge boson $B'$ enters into the covariant derivative by $ D_{\mu}=\partial_{\mu} -igW_{\mu}^{a}T_{a} - ig'\frac{Y}{2}B_{\mu}-ig_{X}B'_{\mu}$ where $g_{X}$ is the $U(1)_{X}$ coupling. Then, gauge boson masses arise when SSB takes place resulting in the following mass matrix for neutral gauge bosons:

\begin{align}
    M_{0}^{2}=\frac{1}{4}\begin{pmatrix}
    g^{2} v^{2} & -gg'v^{2} & -\frac{2}{3}g g_{X} v^{2}(1+\cos^{2}\beta) \\
    * & g'{}^{2} v^{2} & \frac{2}{3}g'g_{X} v^{2}(1+\cos^{2}\beta) \\
    * & * & \frac{4}{9}g_{X}^{2} \left[v_{\chi}^{2}+(1+3\cos^{2}\beta)v^2\right]
    \end{pmatrix}, \nonumber
\end{align}
where we have defined:
\begin{align}
    v^{2}&=v_{1}^{2}+v_{2}^{2}&
    \tan\beta&= \frac{v_{1}}{v_{2}} 
\end{align}

while for charged gauge bosons a SM like relationship $M_{W}=\frac{gv}{2}$ is obtained after the transformation $W^{\pm}_{\mu}=(W_{\mu}^{1}\mp W_{\mu}^{2})/\sqrt{2}$ which gives us an important restriction among VEVs:
\begin{align}
    v^{2}&=v_{1}^{2}+v_{2}^{2}=246^{2}\; \text{GeV}^{2}.
\end{align}
Mass eigenstates represents the photon, $Z$ boson and an additional $Z'$ neutral gauge boson whose masses are given by:
\begin{align}
    M_{\gamma}&=0 & M_{Z}&=\frac{gv}{2\cos\theta_{W}} &
    M_{Z'}&\approx \frac{g_{X}v_{\chi}}{3}
\end{align}
where $\tan\theta_{W}=\frac{g'}{g}$ is the Weinberg angle and the rotation between flavor and mass eigenstates is given by:
\begin{equation}
\begin{pmatrix}
A_{\mu}\\ Z_{\mu}\\Z'_{\mu}
\end{pmatrix}=
\begin{pmatrix}
\sin\theta_W & \cos\theta_W & 0\\
\cos\theta_W\cos\theta_Z & -\sin\theta_W\cos\theta_Z & \sin\theta_Z\\
-\cos\theta_W\sin\theta_Z & \sin\theta_W\sin\theta_Z & \cos\theta_Z\\
\end{pmatrix}
\begin{pmatrix}
W_{\mu}^{3}\\ B_{\mu}\\ B'_{\mu}
\end{pmatrix},    
\end{equation}
where $\theta_Z$ is a small mixing angle between the $Z$ and $Z'$ bosons which decouples in the $\theta_{Z} \rightarrow 0$ limit given by equation \ref{thetaz}. It is worth to notice that the zero entry implies that there is no coupling between $Z$ and $Z'$ particle. The mixing angle is defined by:
\begin{equation}\label{thetaz}
\sin\theta_Z=(1+\cos^2\beta)\frac{2g_X \cos\theta_W}{3g}\left(\frac{M_Z}{M_{Z'}}\right)^2.
\end{equation}

\subsection{Scalar Masses}

The most general scalar potential in agreement with the new symemtry is given by:
\begin{align}
V &= \mu_{1}^{2}\phi_{1}^{\dagger}\phi_{1} + \mu_{2}^{2}\phi_{2}^{\dagger}\phi_{2} + \mu_{\chi}^{2}\chi^{*}\chi + \mu_{\sigma}^{2}\sigma^{*}\sigma + \frac{f}{\sqrt{2}}\left(\phi_{1}^{\dagger}\phi_{2}\chi ^{*} + \mathrm{h.c.} \right) + \frac{f'}{\sqrt{2}}\left(\phi_{1}^{\dagger}\phi_{2}\sigma ^{*} + \mathrm{h.c.} \right) \nonumber\\ 
& + \lambda_{1}\left(\phi_{1}^{\dagger}\phi_{1}\right)^{2} + \lambda_{2}\left(\phi_{2}^{\dagger}\phi_{2}\right)^{2} 
 + \lambda_{3}\left(\chi^{*}\chi \right)^{2} + \lambda_{4}\left(\sigma^{*}\sigma \right)^{2} + \lambda_{5}\left(\phi_{1}^{\dagger}\phi_{1}\right) \left(\phi_{2}^{\dagger}\phi_{2}\right)
 + \lambda'_{5}\left(\phi_{1}^{\dagger}\phi_{2}\right)\left(\phi_{2}^{\dagger}\phi_{1}\right)	\nonumber\\ 
 & + \left(\phi_{1}^{\dagger}\phi_{1}\right)\left[ \lambda_{6}\left(\chi^{*}\chi \right) + \lambda'_{6}\left(\sigma^{*}\sigma \right) %+ \lambda''_{6}\left(\chi^{*}\sigma+\mathrm{h.c.} \right) 
 \right] + \left(\phi_{2}^{\dagger}\phi_{2}\right)\left[ \lambda_{7}\left(\chi^{*}\chi \right) + \lambda'_{7}\left(\sigma^{*}\sigma \right) %+ \lambda''_{7}\left(\chi^{*}\sigma+\mathrm{h.c.} \right) 
 \right]  + \lambda_{8}\left(\chi^{*}\chi \right)\left(\sigma^{*}\sigma \right) \nonumber\\	
 &+ \lambda'_{8}\left[\left(\chi^{*}\sigma \right)\left(\chi^{*}\sigma \right) + \mathrm{h.c.} \right].
\end{align}

This potential provides the following mass matrices for charged, CP-even and CP-odd scalars after SSM takes place. First, charged scalar bosons mass matrix is written in the basis $(\phi^{\pm}_{1},\phi^{\pm}_{2})$, whose rotation matrix is given by $R_{C}$

\begin{align}
\mathit{M}_{\mathrm{C}}^{2} &= \frac{1}{4}
\begin{pmatrix}
	-f\dfrac{v_{\chi}v_{2}}{v_{1}}-\lambda_{5}'{{v_{2}}^{2}} & 
	 fv_{\chi}+\lambda_{5}'v_{1}v_{2}		\\
	 fv_{\chi}+\lambda_{5}'v_{1}v_{2}		&
	-f\dfrac{v_{\chi}v_{1}}{v_{2}}-\lambda_{5}'{{v_{1}}^{2}}
\end{pmatrix}
& \mathit{R}_{\mathrm{C}} &= 
\begin{pmatrix}
c_{\beta}	&	s_{\beta}	\\	-s_{\beta}	&	c_{\beta}
\end{pmatrix},
\end{align}

where the masses are given by:
\begin{align}
m_{G_{W}^{\pm}}^{2} &= 0 & m_{H^{\pm}}^{2} &= -\frac{1}{4}\frac{f v_{\chi}}{s_{\beta}c_{\beta}} -\frac{1}{4}\lambda_{5}' v^2,
\end{align}

As expected from Goldstone theorem, there is a massless boson responsible of the $W_{\mu}^{\pm}$ mass and a heavy charged scalar $H^{\pm}$ whose mass is dominated by the $v_{\chi}$ VEV. 

In the case of CP-odd scalars, mass matrix is written  in the basis $(\eta_{1},\eta_{2},\zeta_{\chi})$ and the rotation matrix is given by $R_{I}$.

\begin{align}
\mathit{M}_{\mathrm{I}}^{2} &= -\frac{f}{4}
\begin{pmatrix}
\dfrac{ {v_{2}}\, {v_{\chi}}}{ {v_{1}}} & - {v_{\chi}} &  {v_{2}}\\
- {v_{\chi}} & \dfrac{ {v_{1}}\, {v_{\chi}}}{ {v_{2}}} & - {v_{1}}\\
 {v_{2}} & - {v_{1}} & \dfrac{ {v_{1}}\, {v_{2}}}{ {v_{\chi}}}
\end{pmatrix} & \mathit{R}_{\mathrm{I}} &= \begin{pmatrix}
c_{\beta}	&	s_{\beta}	&	0	\\
-s_{\beta}	&	c_{\beta}	&	0	\\
0	&	0	&	1
\end{pmatrix}
\begin{pmatrix}
c_{\gamma}	&	0	&	s_{\gamma}	\\
0	&	 1	&	0	\\
-s_{\gamma}	&	0	&	 c_{\gamma}
\end{pmatrix},
\end{align}
where the masses are given by:
\begin{align}
m_{G_{Z}^{0}}^{2}&=	0 & m_{G_{Z'}^{0}}^{2}&=0 & m_{A^{0}}^{2}&=	-\frac{1}{4}\frac{f v_{\chi}}{s_{\beta}c_{\beta}s_{\gamma}^{2}},
\end{align}
and $\gamma$ describes the doublet-singlet mixing $\tan{\gamma}= {v_{\chi}}/{v s_{\beta}c_{\beta}}$. There are two massless scalars as expected from the $Z$ and $Z'$ masses and just like the charged scalar, there is an additional heavy pseudoscalar particle $A^{0}$ in the $v_{\chi}$ scale. 

Finally, regarding CP-even scalars the following mass matrix is obtained in the basis $(h_{1}, h_{2}, \xi_{\chi})$:

\begin{equation}
\label{eq:Mass-matrix-scalar-real}
\mathit{M}_{\mathrm{R}}^{2} = 
\begin{pmatrix}
	\lambda_{1} {v_{1}^{2}}-\dfrac{1}{4}\dfrac{f v_{\chi} v_{2}}{v_{1}} &
	\hat{\lambda}_{5} { v_{1} v_{2}}+\dfrac{1}{4}{f v_{\chi}} &
	\dfrac{1}{4}\lambda_{6}{ v_{1} v_{\chi}}+\dfrac{1}{4}{f v_{2}}		\\
	\hat{\lambda}_{5} { v_{1} v_{2}}+\dfrac{1}{4}{f v_{\chi}} &
	\lambda_{2} {v_{2}^{2}}-\dfrac{1}{4}\dfrac{f v_{\chi} v_{1}}{v_{2}} & 
	\dfrac{1}{4}\lambda_{7}{ v_{2} v_{\chi}}+\dfrac{1}{4}{f v_{1}}		\\
	\dfrac{1}{4}\lambda_{6}{ v_{1} v_{\chi}}+\dfrac{1}{4}{f v_{2}}	& 
	\dfrac{1}{4}\lambda_{7}{ v_{2} v_{\chi}}+\dfrac{1}{4}{f v_{1}} & 
	\lambda_{3} {v_{\chi}^{2}}-\dfrac{1}{4}\dfrac{f v_{1} v_{2}}{v_{\chi}}
\end{pmatrix},
\end{equation}
where $\hat{\lambda}_{5}=\left(\lambda_{5}+\lambda'_{5}\right)/2$. To handle this mass matrix, first we consider a numerical exploration with all $\lambda_{i}$ couplings with a random value between 0 and 1 which showed that $v_{\chi}\sim 10^{7}$ GeV in order to have all heavy particles above $1$ TeV. Then, we can implement a seesaw-like mechanism through the condition $ |f|\upsilon _{\chi}, \upsilon _{\chi} ^2 \gg \upsilon ^2 $ in the matrix elements. Consequently, the heaviest particle mass can be written in a good approximation as the $3\times 3$ mass matrix element. We define the following blocks so we can write the rotated $2\times 2$ matrix in Eq. (\ref{mh2}).
\begin{eqnarray}
{\mathcal{M}_{1}} &=& 
\begin{pmatrix}
	\lambda_{1} {v_{1}^{2}}-\dfrac{1}{4}\dfrac{f v_{\chi} v_{2}}{v_{1}} &
	\hat{\lambda}_{5} { v_{1} v_{2}}+\dfrac{1}{4}{f v_{\chi}}\\
	\hat{\lambda}_{5} { v_{1} v_{2}}+\dfrac{1}{4}{f v_{\chi}} &
	\lambda_{2} {v_{2}^{2}}-\dfrac{1}{4}\dfrac{f v_{\chi} v_{1}}{v_{2}}
\end{pmatrix}, \nonumber	\\
{\mathcal{M}_{12}^{\mathrm{T}}}& =& 
\begin{pmatrix}
	\dfrac{\lambda_{6}{ v_{1} v_{\chi}}}{4}+\dfrac{{f v_{2}}}{4}		\\
	\dfrac{\lambda_{7}{ v_{2} v_{\chi}}}{4}+\dfrac{{f v_{1}}}{4}
\end{pmatrix}
\approx\begin{pmatrix}
	\dfrac{\lambda_{6}{ v_{1} v_{\chi}}}{4}	\\
	\dfrac{\lambda_{7}{ v_{2} v_{\chi}}}{4}
\end{pmatrix},\nonumber	\\
{\mathcal{M}_{2}} &=& \lambda_{3} {v_{\chi}^{2}}-\dfrac{1}{4}\dfrac{f v_{1} v_{2}}{v_{\chi}}
\approx \lambda_{3} {v_{\chi}^{2}}.
\end{eqnarray}

so the rotated $2\times 2$ matrix containing the light scalar is written as:
\begin{align}
M^{2}_{hH} &\approx {\mathcal{M}_{1}} - {\mathcal{M}_{12}^{\mathrm{T}}} {\mathcal{M}_{2}}^{-1} {\mathcal{M}_{12}^{\mathrm{T}}} \nonumber\\
&\approx
    \left(
\begin{array}{cc}
 \left(\lambda _1-\frac{\lambda _6^2}{16 \lambda _3}\right) v_1^2-\frac{f v_2 v_{\chi }}{4 v_1} &  \frac{f v_{\chi }}{16}+\left( \lambda _5-\frac{\lambda _6 \lambda _7}{16\lambda _3}\right) v_1 v_2\\
 \frac{f v_{\chi }}{16}+\left( \lambda _5-\frac{\lambda _6 \lambda _7}{16\lambda _3}\right) v_1 v_2 & \left(\lambda _2-\frac{\lambda _7^2}{16\lambda _3}\right) v_2^2-\frac{ f v_1 v_{\chi }}{4 v_2} \\
\end{array}
\right)\label{mh2}
\end{align}
where we have neglected electroweak additive terms. Finally, after some algebra an neglecting small contributions, the mass eigenvalues can be written as:

\begin{align}
    m_{{H}_{\chi}}^{2}&=\lambda_{3}v_{\chi}^{2}, & m_{H}^{2} &\approx -\frac{1}{2}\frac{f v_{\chi}}{s_{\beta}c_{\beta}}, & m_{h}^{2} &\approx  v^{2}(s_{\beta}^{4}\gamma_{1}  + c_{\beta}^{4}\gamma_{2}+ s_{\beta}^{2}c_{\beta}^{2}\gamma_{3}),
\end{align}
where
\begin{align}
    \gamma_{1}&=2\lambda_{1}-\frac{\lambda_{6}^{2}}{8\lambda_{3}},& \gamma_{2}&=2\lambda_{2}-\frac{\lambda_{7}^{2}}{8\lambda_{3}}, & \gamma_{3}&=4\lambda_{5}-\frac{\lambda_{6}\lambda_{7}}{4\lambda_{3}},
\end{align}
 where the rotation matrix is given by:
 \begin{align}
     R_{S}&=\begin{pmatrix}
     c_{12} & -s_{12} & 0 \\
     s_{12} & c_{12} & 0 \\
     0 & 0 & 1
     \end{pmatrix}
     \begin{pmatrix}
     1 & 0 & \frac{\lambda _6 v_1}{4 \lambda _3 v_{\chi }} \\
     0 & 1 & \frac{\lambda _7 v_2}{4 \lambda _3 v_{\chi }} \\
     -\frac{\lambda _6 v_1}{4 \lambda _3 v_{\chi }} & -\frac{\lambda _7 v_2}{4 \lambda _3 v_{\chi }} & 1
     \end{pmatrix} \nonumber\\
     &=\begin{pmatrix}
         c_{12} & -s_{12} & c_{12}\frac{\lambda _6 v_1}{4 \lambda _3 v_{\chi }} - s_{12}\frac{\lambda _7 v_2}{4 \lambda _3 v_{\chi }}\\
         s_{12} & c_{12} & s_{12}\frac{\lambda _6 v_1}{4 \lambda _3 v_{\chi }} + c_{12}\frac{\lambda _7 v_2}{4 \lambda _3 v_{\chi }} \\
          -\frac{\lambda _6 v_1}{4 \lambda _3 v_{\chi }} & -\frac{\lambda _7 v_2}{4 \lambda _3 v_{\chi }} & 1
     \end{pmatrix},
 \end{align}
where
\begin{align}
    s_{12}&\approx -\frac{1}{\sqrt{1+\left(\frac{ 4 \lambda _3 \left(f v_1 v_{\chi }+2 m_{h}^2 v_2-4 \lambda _2 v_2^3\right)+\lambda _7^2 v_2^3}{v_2 \left(4 f \lambda _3 v_{\chi }+\left(16 \lambda _3 \lambda _5-\lambda _6 \lambda _7\right) v_1 v_2\right)}\right)^{2}}} \approx -\frac{v_{2}}{v}
\end{align}

being the first rotation associated with the seesaw rotation. the lightest scalar can be identified with the SM Higgs boson and it can be noticed that it does not depend on the $U(1)_X$ symmetry breaking scale. 
\subsection{Fermion Masses}
Since we are interested in the muon anomalous magnetic moment, lets consider only lepton masses whose most general interaction lagrangian reads, for charged and neutral leptons, as follows:

\begin{equation}
\begin{split}
-\mathcal{L}_{Y,E} &= 
h_{2e}^{e\mu} \overline{\ell^{e}_{L}}\phi_{2}e^{\mu}_{R} + h_{2e}^{\mu\mu} \overline{\ell^{\mu}_{L}}\phi_{2}e^{\mu}_{R} + 
h_{2e}^{\tau e}\overline{\ell^{\tau}_{L}}\phi_{2}e^{e}_{R} + h_{2e}^{\tau\tau}\overline{\ell^{\tau}_{L}}\phi_{2}e^{\tau}_{R} +	
h_{1e}^{E}\overline{\ell^{e}_{L}}\phi_{1}E_{R} + h_{1\mu}^{E}\overline{\ell^{\mu}_{L}}\phi_{1}{E}_{R} \\ &+
g_{\chi' E}\overline{E_{L}}\chi E_{R} + g_{\chi \mathcal{E}}\overline{\mathcal{E}_{L}}\chi^{*} \mathcal{E}_{R} -\mu_{E}\bar{E}_{L}\mathcal{E}_{R} -\mu_{\mathcal{E}}\bar{\mathcal{E}}_{L}E_{R} +
h_{\sigma e}^{E}\overline{E_{L}}\sigma e^{e}_{R} + h_{\sigma \mu}^{\mathcal{E}}\overline{\mathcal{E}_{L}}\sigma^{*} e^{\mu}_{R} + 
h_{\sigma \tau}^{E}\overline{E_{L}}\sigma e^{\tau}_{R}  + \mathrm{h.c.}
\end{split}
\label{eq:Electron-Lagrangian}
\end{equation}

\begin{equation}\label{LN}
\begin{split}
-\mathcal{L}_{Y,N} &= 
h_{2e}^{\nu e}\overline{\ell^{e}_{L}}\tilde{\phi}_{2}\nu^{e}_{R} + 
h_{2e}^{\nu \mu}\overline{\ell^{e}_{L}}\tilde{\phi}_{2}\nu^{\mu}_{R} + 
h_{2e}^{\nu \tau}\overline{\ell^{e}_{L}}\tilde{\phi}_{2}\nu^{\tau}_{R} + 
h_{2\mu}^{\nu e}\overline{\ell^{\mu}_{L}}\tilde{\phi}_{2}\nu^{e}_{R} +
h_{2\mu}^{\nu \mu}\overline{\ell^{\mu}_{L}}\tilde{\phi}_{2}\nu^{\mu}_{R} + 
h_{2\mu}^{\nu \tau}\overline{\ell^{\mu}_{L}}\tilde{\phi}_{2}\nu^{\tau}_{R} \\ &+
h_{\chi i}^{\nu j} \overline{\nu_{R}^{i\;C}} \chi^{*} N_{R} +
\frac{1}{2} \overline{N_{R}^{i\;C}} M^{ij}_{N} N_{R}^{j}  + \mathrm{h.c.},
\end{split}
\end{equation}

The above lagrangian generates a specific zero-texture for mass matrices which can be addressed to the fermion mass hierarchy problem and might ensure PMNS matrix reproducibilty. It also has been widely studied in \cite{textures} as well as their possible origin.

\subsubsection*{Charged leptons}
The mass matrix is written in the flavor basis $(e^{e},e^{\mu} , e^{\tau}, E, \mathcal{E})$:

\begin{align}
    \mathcal{M}_{E}&=\frac{1}{\sqrt{2}}\left(\begin{array}{ c c c |c c}\\
    v_{2}\Sigma_{11}                           & h_{2e}^{e\mu}v_{2}     & v_{2}\Sigma_{13} &  h_{1e}^{E}v_{1}    & 0 \\
    0                           & h_{2e}^{\mu\mu}v_{2}   & 0 &  h_{1\mu}^{E}v_{1}  & 0 \\
    h_{2e}^{\tau e}v_{2}  & 0                            & h_{2e}^{\tau\tau}v_{2} & 0 & 0 \\ \hline
    0 & 0 & 0 & {g}_{\chi E}v_{\chi} & -\mu_{E} \\
    0 & 0 & 0 & -\mu_{\mathcal{E}} & g_{\chi\mathcal{E}}v_{\chi}  \\
    \end{array} \right)
\end{align}

\begin{figure}[H] 
\centering
\includegraphics[scale=0.4]{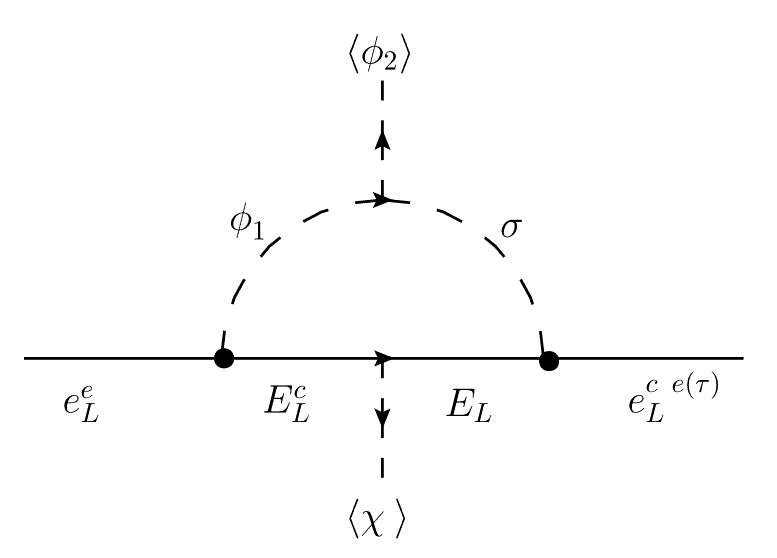}\vspace{-0.5cm}
\caption{\small Charged leptons mass one-loop correction.}
\label{fig-oneloop-2}
\end{figure}

The $\mathcal{E}$ lepton is decoupled decoupled from SM leptons so a rotation can be done to decouple one exotic mass eigenstates represented by the angle $\theta_{E\mathcal{E}}^{L/R}$.
It turns out that the squared matrix $\mathcal{M}_{E}\mathcal{M}_{E}^{\dagger}$ has null determinant so the electron is massless at tree level. However, it can acquire a small mass at one-loop level by considering the correction shown in  figure \ref{fig-oneloop-2}, which adds the following two terms:
\begin{align}
    \Delta \mathcal{L}&=\frac{v_{2}}{2}\Sigma_{11}e^{e}_{L}e^{e}_{R} + \frac{v_{2}}{2}\Sigma_{13}e^{e}_{L}e^{\tau}_{R} .
\end{align}

To obtain mass eigenvalues we consider the squared mass matrix $\mathcal{M}_{E}\mathcal{M}_{E}^{\dagger}$ which is diagonalized with the rotation of left handed leptons. Diagonalization can be done using first a seesaw rotation to decouple the exotic lepton. As a result, $\tau$ lepton decouples at tree level. However, a second seesaw rotation is done and the resulting $2\times 2$ matrix is diagonalized, resulting in the following mass eigenvalues:

\begin{align}
m_{e}^{2}&=\frac{1}{2}v_{2}^{2}v_{2}^{2}\frac{t_{3}^{2}}{2m_{\tau}^{2}}, & m_{\mu}^{2}&=\frac{1}{2}v_2^{2}\left[(h_{2e}^{e\mu})^2+(h_{2e}^{\mu\mu})^2\right], \\
m_{\tau}^{2}&=\frac{1}{2}v_2^2\left[(h_{2e}^{\tau e})^2+(h_{2e}^{\tau \tau})^2\right], & m_{E}^{2}&=\frac{1}{2}g_{\chi' E}^2\; v_\chi^2, \\
m_{\mathcal{E}}^{2}&=\frac{1}{2}g_{\chi \mathcal{E}}^2 v_\chi^2,
 \end{align}
where the rotation matrix is given by $V^{L}=V_{3}^{\ell}V_{2}^{\ell}V_{1}^{\ell}$ being $V_{1}^{\ell}$ and $V_{2}^{\ell}$ related to seesaw rotations
\begin{align}\label{v1l}
  V_{1}^{\ell}&=\begin{pmatrix}
      1 & 0 & 0 & -\frac{g_{\chi \mathcal{E} } h_e^e v_{\chi } v'_1}{g_{\text{$\chi $E}} g_{\chi \mathcal{E} } v_{\chi } v'_{\chi }-\mu_{E} \mu _{\mathcal{E} }} & \frac{h_e^e \mu_{E} v'_1}{\mu_{E} \mu _{\mathcal{E} }-g_{\text{$\chi $E}} g_{\chi \mathcal{E} } v_{\chi } v'_{\chi }} \\
      0 & 1 & 0 & - \frac{g_{\chi \mathcal{E} } h_{\mu }^e v_{\chi } v'_1}{g_{\text{$\chi $E}} g_{\chi \mathcal{E} } v_{\chi } v'_{\chi }-\mu_{E} \mu _{\mathcal{E} }} & \frac{h_{\mu }^e \mu_{E} v'_1}{\mu_{E} \mu _{\mathcal{E} }-g_{\text{$\chi $E}} g_{\chi \mathcal{E} } v_{\chi } v'_{\chi }} \\
      0 & 0 & 1 & 0 &0 \\
      \frac{g_{\chi \mathcal{E} } h_e^e v_{\chi } v'_1}{g_{\text{$\chi $E}} g_{\chi \mathcal{E} } v_{\chi } v'_{\chi }-\mu_{E} \mu _{\mathcal{E} }}  &  \frac{g_{\chi \mathcal{E} } h_{\mu }^e v_{\chi } v'_1}{g_{\text{$\chi $E}} g_{\chi \mathcal{E} } v_{\chi } v'_{\chi }-\mu_{E} \mu _{\mathcal{E} }} & 0 & 1 &0 \\
       -\frac{h_e^e \mu_{E} v'_1}{\mu_{E} \mu _{\mathcal{E} }-g_{\text{$\chi $E}} g_{\chi \mathcal{E} } v_{\chi } v'_{\chi }} & -\frac{h_{\mu }^e \mu_{E} v'_1}{\mu_{E} \mu _{\mathcal{E} }-g_{\text{$\chi $E}} g_{\chi \mathcal{E} } v_{\chi } v'_{\chi }} & 0 & 0 &1 
    \end{pmatrix} ,
    \end{align}
    \begin{align}
    V_{2}^{\ell}&=\begin{pmatrix}
      1 & 0 & - \frac{m_{e}^{2}}{t_{3}v_{2}v_{2}^{\prime}} &0 & 0\\
      0 & 1 & 0 &0 & 0 \\
       \frac{m_{e}^{2}}{t_{3}v_{2}v_{2}^{\prime}} & 0 & 1 & 0 &0 \\
      0 & 0 & 0 & \cos\theta_{E\mathcal{E}}^{L} & -\sin\theta_{E\mathcal{E}}^{L} \\
      0 & 0 & 0 & \sin\theta_{E\mathcal{E}}^{L} & \cos\theta_{E\mathcal{E}}^{L} 
    \end{pmatrix}, & V_{3}^{\ell}&= \begin{pmatrix}
      \cos\theta_{e\mu} & \sin\theta_{e\mu} &0 &0 & 0\\
      -\sin\theta_{e\mu} & \cos\theta_{e\mu} &0 & 0 &0 \\
      0 & 0 & 1 & 0 & 0\\
      0 & 0 & 0 & 1 & 0 \\
      0 & 0 & 0 & 0 & 1
    \end{pmatrix} 
    \end{align}
    where
    \begin{align}
    t_{3}&=\Sigma_{11}h_{2 e}^{\tau e}+\Sigma_{13} h_{2 e}^{\tau \tau },  &  \sin\theta_{e\mu}&=-\frac{2 m_{\mu}^{2}-v_2^{\prime 2} (h_{2 e}^{\mu\mu})^{2} }{ h_{2 e}^{e\mu } h_{2 e}^{\mu\mu} v_2^{\prime 2}\sqrt{\left(\frac{v_2^{\prime 2} (h_{2 e}^{\mu\mu})^{2} -2 m_{\mu}^{2}}{h_{2 e}^{e\mu  }  h_{2 e}^{\mu\mu } v_2^{\prime 2}}\right)^2+1}}. \label{v3l}
\end{align}
Similarly, when diagonalizing the squared matrix $\mathcal{M}_{E}^{\dagger}\mathcal{M}_{E}$ we get the right handed rotation $V^{R}=U_{2}^{\ell}U_{1}^{\ell}$ which can be written as :

\begin{align}
    U_{2}^{\ell}&=\begin{pmatrix}
 \cos\theta_{e\tau} & \frac{v_{2}^{2}h_{2 e}^{e\mu} \left(\sin\theta_{e\tau} \Sigma_{13}-\cos\theta_{e\tau} \Sigma_{11}\right) }{2 m_{\mu}^{2}} & -\sin\theta_{e\tau} & 0 & 0 \\
 \frac{\Sigma_{11} v_{2}^{2} h_{2 e}^{e\mu}}{2 m_{\mu}^{2}} & 1 & \frac{\Sigma_{13} v_{2}^{2} h_{2 e}^{e\mu}}{2 m_{\mu}^{2}} & 0 & 0 \\
 \sin\theta_{e\tau} & -\frac{v_{2}^{2} h_{2 e}^{e\mu} \left(\cos_{theta_{e\tau}}\Sigma_{13}+\sin\theta_{e\tau} \Sigma_{11}\right) }{2 m_{\mu}^{2}} & \cos\theta_{e\tau} & 0 & 0 \\
 0 & 0 & 0 & \cos\theta_{E\mathcal{E}}^{R} & -\sin\theta_{E\mathcal{E}}^{R} \\
 0 & 0 & 0 & \sin\theta_{E\mathcal{E}}^{R} & \cos\theta_{E\mathcal{E}}^{R} \\
\end{pmatrix}, &
U_{1}^{\ell}&=\begin{pmatrix}
\mathcal{I}_{3\times 3} & -\Theta^{T}  \\
\Theta & \mathcal{I}_{2\times 2}
\end{pmatrix}
\end{align}
 with
\begin{align}
    \Theta=\begin{pmatrix}
        \frac{\Sigma_{11} v_{2}v_{1} h_{1e}^{E} \left(g_{\chi \mathcal{E} }^2 v_{\chi }^2+\mu_{E}^{2}\right)}{4 m_{E}^2 m_{\mathcal{E}}^2} & \frac{v_{1}v_{2} \left(g_{\chi \mathcal{E} }^2 v_{\chi }^2+\mu_E^2\right) \left(h_{1e}^{E} h_{2e}^{e\mu}+h_{\mu }^{E} h_{2e}^{\mu \mu }\right)}{4 m_{E}^2 m_{\mathcal{E}}^2} & \frac{\Sigma_{13} v_{1}v_{2} h_{1e}^{E} \left(g_{\chi \mathcal{E} }^2 v_{\chi }^2+\mu_E^2\right)}{4 m_{E}^2 m_{\mathcal{E}}^2}  \\
\frac{\Sigma_{11} v_{2}v_{1} h_{1e}^{E} \left(\mu_{E} g_{\chi E} v_{\chi}+g_{\chi \mathcal{E} } v_{\chi } \mu _{\mathcal{E}}\right)}{4 m_{E}^2 m_{\mathcal{E}}^2}  & \frac{v_{1}v_{2} \left(h_{1e}^{E} h_{2 e}^{e\mu}+h_{1\mu}^{E} h_{2 e}^{\mu \mu }\right) \left(\mu_{E} g_{\chi E} v_{\chi}+g_{\chi \mathcal{E} } v_{\chi } \mu_{\mathcal{E}}\right)}{4 m_{E}^2 m_{\mathcal{E}}^2} & \frac{\Sigma_{13} v_{1}v_{2} h_{1e}^E \left(\mu_E g_{\chi E} v_{\chi}+g_{\chi \mathcal{E} } v_{\chi } \mu_{\mathcal{E} }\right)}{4 m_{E}^2 m_{\mathcal{E}}^2} 
    \end{pmatrix}.
\end{align}

\subsubsection*{Neutral leptons}
In this case, contrary to SM theory neutrino are massive giving rise to the neutrino oscillation phenomena \cite{nuoscillations} which has motivated many experiments who have confirmed it and having the measurement of their mass as one of their major goals \cite{nuexperiments}. In this scenario, there are right-handed and Majorana neutrinos which provides the SM neutrinos a finite mass value via inverse seesaw mechanism. According to the general Yukawa lagrangian show in Eq. (\ref{LN})  the neutrino mass matrix, written in the basis $\left(\begin{matrix}{\nu^{e,\mu,\tau}_{L}},\,\left(\nu^{e,\mu,\tau}_{R}\right)^{C},\,\left(N^{e,\mu,\tau}_{R}\right)^{C}\end{matrix}\right)^{\mathrm{T}}$, is given as follows:

\begin{align}
\mathcal{M}_{\nu} &=\begin{pmatrix}
    0 & m_{D} & 0 \\
    m_{D}^{T} & 0 & M_{D} \\
    0 & M_{D}^{T} & M_{M}
    \end{pmatrix},     
\end{align}

where the block matrices are given by:
\begin{align}
 &m_{D}=\frac{v_{2}}{\sqrt{2}}\begin{pmatrix}
    h_{2e}^{\nu e} & h_{2e}^{\nu \mu} & h_{2e}^{\nu \tau} \\
    h_{2\mu}^{\nu e} & h_{2\mu}^{\nu \mu} & h_{2\mu}^{\nu \tau} \\
    0 & 0 & 0
    \end{pmatrix},\ \ \
     (M_{D})^{ij}=\frac{v'_{\chi}}{\sqrt{2}}({h}_{\chi}^{\prime \nu})^{ij}, \ \ \ \ \  (M_{M})_{ij}=\frac{1}{2}M_{ij}.
\end{align}

If we assume a hierachy among parameters, such as $M_{M}\ll m_{D} \ll M_{D} $ the matrix is diagonalized via an inverse seesaw mechanism (ISS) (see appendix \ref{appendinx} for further details). Consequently, block diagonalization is done by a rotation matrix $\mathbb{V}_{SS}$ given by:

\begin{align}
\mathbb{V}_{SS}\mathcal{M}_{\nu}\mathbb{V}_{SS}^{\dagger}&\approx\begin{pmatrix}
m_{light}&0\\
0&m_{heavy}
\end{pmatrix}, &  \\
\mathbb{V}_{SS}&=\begin{pmatrix}
I&-\Theta_{\nu}\\
\Theta_{\nu}^{T}&I
\end{pmatrix}, & \Theta_{\nu}&=\begin{pmatrix}
0&M_{D}^{T}\\
M_{D}&M_{M}
\end{pmatrix}^{-1}\begin{pmatrix}
m_{D}^{T}\\0
\end{pmatrix},
\end{align}

where $m_{light}=m_{D}^{T}(M_{D}^{T})^{-1}M_{M}(M_{D})^{-1}m_{D}$ is the $3\times3$ mass matrix containing the SM light neutrinos and encodes the information of the PMNS matrix while $m_{heavy}$ matrix mixes right handed and Majorana neutrino eigenstates, which is given by:
\begin{align}\label{nuheavy}
m_{heavy}\approx\begin{pmatrix}0&M_{D}^{T}\\
M_{D}&M_{M}
\end{pmatrix}  .  
\end{align}
For simplicity and considering too heavy exotic neutrinos to be observed and nearly indistinguishable for us, we can take the particular case where  $M_{D}$ is diagonal and $M_{M}$ is proportional to the identity to explore one of the possible scenarios of the model.

\begin{align}
M_{D} &= \frac{v_{\chi}}{\sqrt{2}} \left( \begin{matrix}
h_{N\chi 1}	&	0	&	0	\\	0	&	h_{N\chi 2}	&	0	\\	0	&	0	&	h_{\chi N 3}
\end{matrix} \right) &
M_{M} &= \mu_{N} \mathbb{I}_{3\times 3}.
\end{align}

In this way, the light neutrino mass matrix takes the form

\begin{equation}\label{mnu}
m_{\mathrm{light}} = \frac{\mu_{N} v_{2}^{2}}{{h_{N\chi 1}}^{2}v_{\chi}^{2}}
\left( 
\begin{matrix}
	\left( h_{2e}^{\nu e}\right)^{2} + \left( h_{2\mu}^{\nu e} \right)^{2} \rho^{2} &
	{h_{2e}^{\nu e}}\,{h_{2e}^{\nu \mu}} + {h_{2\mu}^{\nu e}}\,{h_{2\mu}^{\nu \mu}}\rho^2 	&
	{h_{2e}^{\nu e}}\,{h_{2e}^{\nu \tau}}+ {h_{2\mu}^{\nu e}}\,{h_{2\mu}^{\nu \tau}}\rho^2 	\\
	{h_{2e}^{\nu e}}\,{h_{2e}^{\nu \mu}} + {h_{2\mu}^{\nu e}}\,{h_{2\mu}^{\nu \mu}}\rho^2	&	
	\left( h_{2e}^{\nu \mu} \right)^{2} + \left( h_{2\mu}^{\nu\mu} \right)^{2} \rho^{2}	&	
	{h_{2e}^{\nu \mu}}\,{h_{2e}^{\nu \tau}}+ {h_{2\mu}^{\nu \mu}}\,{h_{2\mu}^{\nu \tau}}\rho^2	\\
	{h_{2e}^{\nu e}}  \,{h_{2e}^{\nu \tau}}+ {h_{2\mu}^{\nu e}}  \,{h_{2\mu}^{\nu \tau}}\rho^2	&	
	{h_{2e}^{\nu \mu}}\,{h_{2e}^{\nu \tau}}+ {h_{2\mu}^{\nu \mu}}\,{h_{2\mu}^{\nu \tau}}\rho^2	&	
	\left( h_{2e}^{\nu \tau} \right)^{2} + \left( h_{2\mu}^{\nu \tau} \right)^{2} \rho^{2}
\end{matrix} \right),
\end{equation}

where $\rho={h_{N\chi 1}}/{h_{N\chi 2}}$. The matrix $m_{\mathrm{light}}$ has zero determinant for every possible choice of $M_{D}$ and $M_{M}$ since $m_{D}$ has null determinant so at least one neutrino is massless. However, since this is a symmetric matrix its diagonalization is done by its singular value decomposition, which diagonalizes the matrix with the positive square root of the eigenvalues of the squared matrix $m_{light}m_{light}^{\dagger}$.

Related to exotic neutrinos, mass eigenstates are labeled as $\mathcal{N}^{k}$, $k=1,...,6.$ whose eigenvalues can be obtained easily from Eq. (\ref{nuheavy}):
\begin{align}
    m_{\mathcal{N}^{1}}&=\frac{1}{2}(\mu_{N}-\sqrt{\mu_{N}^{2}+2h_{N_{\chi 1}}v_{\chi}^{2}}) & m_{\mathcal{N}^{2}}&=\frac{1}{2}(\mu_{N}-\sqrt{\mu_{N}^{2}+2h_{N_{\chi 2}}v_{\chi}^{2}}) \\
    m_{\mathcal{N}^{3}}&=\frac{1}{2}(\mu_{N}+\sqrt{\mu_{N}^{2}+2h_{N_{\chi 1}}v_{\chi}^{2}}) & m_{\mathcal{N}^{4}}&=\frac{1}{2}(\mu_{N}+\sqrt{\mu_{N}^{2}+2h_{N_{\chi 2}}v_{\chi}^{2}}) \\
    m_{\mathcal{N}^{5}}&=\frac{1}{2}(\mu_{N}-\sqrt{\mu_{N}^{2}+2h_{N_{\chi 3}}v_{\chi}^{2}}) & m_{\mathcal{N}^{6}}&=\frac{1}{2}(\mu_{N}+\sqrt{\mu_{N}^{2}+2h_{N_{\chi 3}}v_{\chi}^{2}})     
\end{align}

\section{Muon $g-2$ contributions}\label{nosusycontributions}
Prior to consider muon anomalous magnetic moment contributions we need to do a parameter fitting that ensures neutrino mass differences and all PMNS physical parameters within the experimental bounds \cite{nufitting}. In our particular case we can consider one particular choice, despite the huge amount of free parameters, where physical masses are correctly reproduced and PMNS matrix as well. On the one hand,  this particular choice involves random values between 0 and 1 for the exotic couplings which lead to a unique determination of the SM leptons parameters. On the other hand, neutrino parameter fitting shows that  $\sqrt{\frac{\mu_{N} v_{2}^{2}}{{h_{N\chi 1}}^{2}v_{\chi}^{2}} }h_{2\alpha}^{\nu\beta} \sim 10^{-6} GeV^{1/2}$ and $\rho \sim 1$ which allows us to consider Yukawa couplings $h_{2\alpha}^{\nu\beta}$ of order 1 and $\frac{\mu_{N} v_{2}^{2}}{{h_{N\chi 1}}^{2}v_{\chi}^{2}} \sim 10^{-12} GeV$ which implies that there are two pairs of degenerate heavy neutrino masses. Furthermore, to recreate the CP violating phase at least one parameter in each column of $m_{D}$ must be complex, which will lead to contributions with a complex phase. However, this allows us to show a general behavior of the model.

We can consider $v_{2}$ to be of order of $\tau$ mass which implies a big $\beta$ mixing angle. To achieve that, one possibility is to we can choose $\gamma_{3}$ to be of order $\sim 10^{3}$ which is always possible if $\lambda_{3}$ is highly suppressed and finally Higgs mass can be guaranteed since $\gamma_{1}$ and $\gamma_{2}$ cause small changes in the scalar mass. The exact parameters used in the following section are:

\begin{align*}
  v_{1} &=245.98 \;GeV & v_{2} &= 3 \;GeV & v_{\chi} &= 10^7 \; GeV & \Sigma_{11} &= -0.0008542 \\ h_{2e}^{e\mu} &= 0.04779 & \Sigma_{13} &= 0.0004265 & 
 h_{2e}^{\mu\mu} &= 0.01400 & h_{2e}^{\tau e} &= 0.005476 \\ 
  h_{2e}^{\tau\tau} &= 0.8375 & h_{1e}^{E} &= 0.5664& h_{1\mu}^{E} &= 0.5119 &   g_{\chi E} &= 0.5527& \\ 
  \mu_{E} &= 0.8296\; GeV &  \mu_{\mathcal{E}} &= 0.0427 \; GeV& g_{\chi\mathcal{E}} &= 0.5974 & Re[h_{e\nu}^{ee}] &= 4.803 \\ 
 Re[h_{2e}^{\nu \mu}] &= 2.016 & Re[h_{2e}^{\nu \tau}] &= 4.722 &  Re[h_{2\mu}^{\nu e}] &= -2.053& Re[h_{2\mu}^{\nu \mu}] &= 1.109 \\  
 Re[h_{2\mu}^{\nu \tau}] &= 1.791 & Im[h_{2e}^{\nu e}] &= -0.2985 & Im[h_{2e}^{\nu \mu}] &= 0.8469 & Im[h_{2e}^{\nu \tau}] &= 0.04768 \\
 h_{N_{\chi 1}} &= 1 & h_{N_{\chi 2}} &= 1 & h_{N_{\chi 3}} &= 1.5 &  \mu_{N} &= 11.11 \;GeV
\end{align*}

where in general the couplings associated with exotic leptons where taken as random numbers between 0 and 1, $\Sigma_{11}$ and $\Sigma_{33}$ where random numbers between 0 and $10^{-4}$. All other parameters have an associated value that reproduces the correct masses and the PMNS matrix.

\subsection{$Z'$ corrections}
\begin{figure}[H]
    \centering
    \includegraphics[scale=0.15]{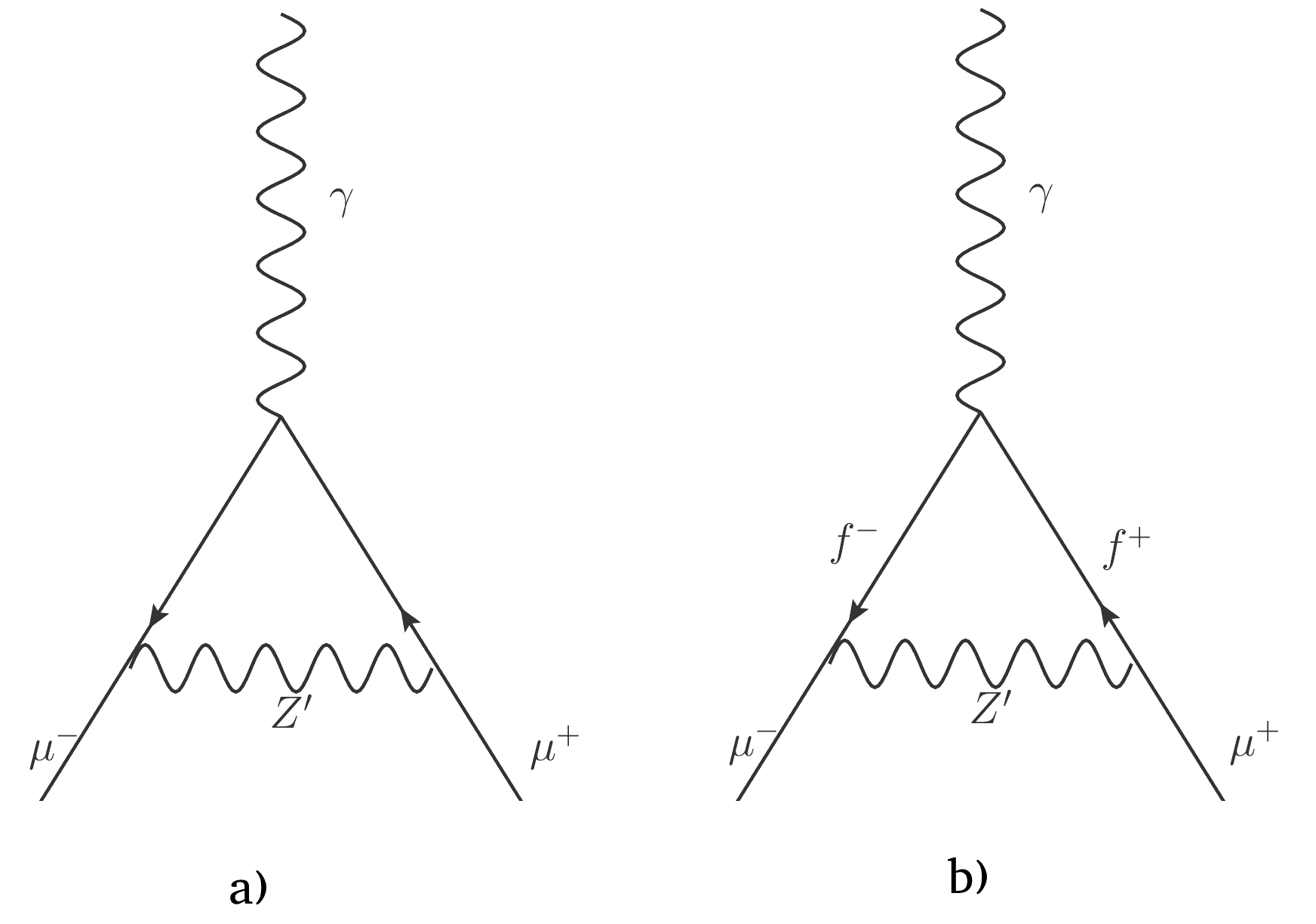}
    \caption{Corrections to muon $g-2$ due to $Z'$ interactions.}
    \label{ZprimeTriangle}
\end{figure}

The interaction lagrangian for the process shown in figure \ref{ZprimeTriangle} includes the $\tau$ lepton and the two exotic singlets, and it is given by:

%\begin{align}
%    \mathcal{L}_{int}(Z')&=-\bar{\mu}Z'_{\mu}\frac{\gamma^{\mu}}{6}\left(P_{L} (3 g's_{\theta_{W}}s_{\theta_{Z}}- 3g_{w}c_{\theta_{W}}s_{\theta_{Z}} +3\frac{g_{X}  \sin^{2}\theta_{e\mu} t_{3} v_{2}^{2} c_{\theta_{Z}}}{m_{\tau}^{2}})+ (6g' s_{\theta_{W}}s_{\theta_{Z}}+2g_{X} c_{\theta_{Z}})P_{R}\right)\mu \nonumber\\
%    &-\bar{\mu}Z'_{\mu}\gamma^{\mu} g_{X}v_{2}^{2}c_{\theta_{Z}} \left( \frac{\sin\theta_{e\mu}t_{3}}{2m_{\tau}^{2}}P_{L}+\frac{\left(\cos\theta_{e\tau} \Sigma_{13}+\sin\theta_{e\tau} \Sigma_{11}\right) h_{2 e}^{e\mu}}{2 m_{\mu}^{2}} P_{R}\right)\tau \nonumber \\
%    &-\bar{\mu}Z'_{\mu}\frac{\gamma^{\mu}}{3} g_{X}c_{\theta_{Z}}\Big(P_{L} (3 V^{L}_{2,3} V^{L}_{4,3}+3 V^{L}_{2,4} V^{L}_{4,4}+2 V^{L}_{2,5} V^{L}_{4,5}) \nonumber \\
%    &+ P_{R} (3 V^{R}_{2,1} V^{R}_{4,1}+3 V^{R}_{2,3} V^{R}_{4,3}+V^{R}_{2,4} V^{R}_{4,4}+2 V^{R}_{2,5} V^{R}_{4,5})\Big)E \nonumber\\
%    &+\bar{\mu}Z'_{\mu}\gamma^{\mu}c_{\theta_{Z}}g_{X} \Bigg( \frac{g_{\chi \mathcal{E}} v_{1} v_{\chi }}{2 m_{E} m_{\mathcal{E}}} \left( s_{\theta_{e\mu}} h_{1e}^E -c_{\theta_{e\mu}} h_{1 \mu }^E  \right)s_{\theta_{E\mathcal{E}}^{L}}P_{L} \nonumber \\
%    &-\frac{v_{1}v_{2} \left(g_{\chi \mathcal{E}}^2 v_{\chi }^2+\mu_E^2\right) \left(h_{1e}^{E} h_{2 e}^{e\mu}+ h_{1\mu }^{E} h_{2 e}^{\mu \mu }\right)}{12 m_{E}^2 m_{\mathcal{E}}^2 }s_{\theta_{E\mathcal{E}}^{R}} P_{R}\Bigg)\mathcal{E}
%\end{align}
\begin{align}
  \mathcal{L}&_{int}(Z')  \approx -\bar{\mu}Z'_{\mu}\frac{\gamma^{\mu}}{3}g_{X}\left(\frac{3  s_{\theta_{e\mu}}^{2} t_{3} v_{2}^{2}}{2m_{\tau}^{2}}P_{L}+ P_{R}\right)\mu -\bar{\mu}Z'_{\mu}\gamma^{\mu} g_{X}v_{2}^{2} \left( \frac{s_{\theta_{e\mu}}t_{3}}{2m_{\tau}^{2}}P_{L}+\frac{\left(c_{\theta_{e\tau}} \Sigma_{13}+s_{\theta_{e\tau}} \Sigma_{11}\right) h_{2 e}^{e\mu}}{2 m_{\mu}^{2}} P_{R}\right)\tau \nonumber \\
    &-\bar{\mu}Z'_{\mu}\gamma^{\mu}g_{X}c_{\theta_{Z}}\Big(P_{L} (V^{L}_{2,3} V^{L}_{4,3}+V^{L}_{2,4} V^{L}_{4,4}+\frac{2}{3}V^{L}_{2,5} V^{L}_{4,5}) + P_{R} (V^{R}_{2,1} V^{R}_{4,1}+V^{R}_{2,3} V^{R}_{4,3}+\frac{1}{3}V^{R}_{2,4} V^{R}_{4,4}+\frac{2}{3} V^{R}_{2,5} V^{R}_{4,5})\Big)E \nonumber\\
    &+\bar{\mu}Z'_{\mu}\gamma^{\mu} \frac{v_{1}c_{\theta_{Z}}g_{X}}{2m_{E}}\Bigg( \sqrt{2}\left( s_{\theta_{e\mu}} h_{1e}^E -c_{\theta_{e\mu}} h_{1 \mu }^E  \right)s_{\theta_{E\mathcal{E}}^{L}}P_{L} -\frac{v_{2}}{3m_{E}} \left(h_{1e}^{E} h_{2 e}^{e\mu}+ h_{1\mu }^{E} h_{2 e}^{\mu \mu }\right)s_{\theta_{E\mathcal{E}}^{R}} P_{R}\Bigg)\mathcal{E}
\end{align}
The rotations involved in the $E$ interaction is not written explicitly because it is highly dependent on the seesaw rotations. In fact, the contribution proportional to $P_{L}$ is or order $\sim 10^{-6}$ and the contribution proportional to $P_{R}$ is of order $\sim 10^{-10}$. Besides, the angle $\theta_{E\mathcal{E}}^{L/R}$ represent the mixing between exotic leptons whether the rotation if of left/right handed leptons and simplification were done by considering $s_{\theta_{Z}}\approx 0$.

The contribution is given by:

\begin{align}
\Delta a_{\mu} (f) &= \frac{1}{8\pi^2}\frac{m_\mu^2}{ M_{Z'}^2 } \int_0^1 dx \frac{g_{v}^2 \ P_{v}(x) + g_{a}^2 \ P_{a} (x) }{(1-x)(1-\lambda^{2} x) +\epsilon^2 \lambda^2 x}\nonumber\\
&\approx  \frac{1}{4\pi^2}\frac{m_\mu^2}{ M_{Z'}^2 }\left\lbrace  g_{v}^2\left[\frac{ M_{f} }{m_{\mu}} -\frac{2}{3}\right] + g_{a}^2 \left[ -\frac{ M_f}{m_{\mu}} -\frac{2}{3}\right] \right\rbrace, \nonumber\\
\label{leptonmuon4}
\end{align}
where $\epsilon = M_{f}/m_{\mu}$ and $\lambda= m_{\mu}/M_{Z'}$, $g_{v}$ and $g_{a}$ the vector and axial couplings respectively and 

\begin{eqnarray}
P_{v}(x) & = &  2x(1-x)(x-2(1-\epsilon))+\lambda^2(1-\epsilon)^2x^2(1+\epsilon-x) \nonumber\\
P_{a}(x) & = &  2x^2(1+x+2\epsilon)+\lambda^2(1+\epsilon)^2x(1-x)(x-\epsilon)
\label{leptonmuon5}
\end{eqnarray}

\begin{figure}[H]
    \centering
    \includegraphics[scale=0.5]{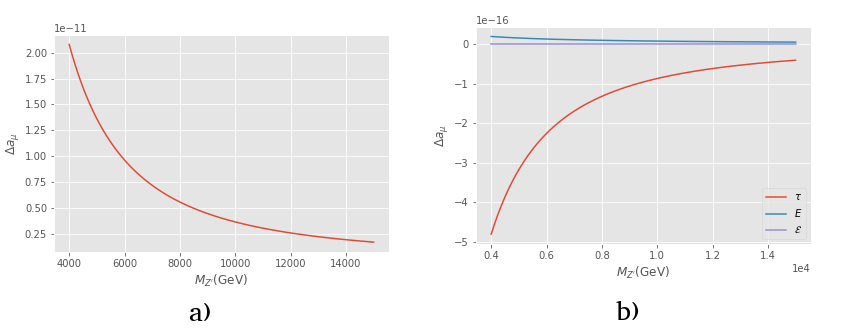}
    \caption{Contribution to $\Delta a$ as a function of $Z'$ mass for in a flavor conserving process(left) and a flavor violating process(right).}
    \label{ZprimeContributions}
\end{figure}

Due to non-universality, we can have in general flavor violation interactions in the model mediated by the $Z'$ gauge boson, however, their contributions are negligible small as shown in figure \ref{ZprimeContributions}

%Flavor Conserving
%$g_{v}=-0.13716$ and $g_{a}=0.137664i$ 
 
%Flavor Violating:
%\begin{align}
%    g_{v}^{\tau}&=0.000193945 & g_{v}^{E}&=7.77656*10^{-7} & g_{v}^{\mathcal{E}}&=0.00132971 \\
%    g_{a}^{\tau}&=0.00021847i & g_{a}^{E}&=7.77519*10^{-7}i & g_{a}^{\mathcal{E}}&=0.00132973 i
%\end{align}

Nevertheless, the contribution of a flavor violating process at a close energy scale is both negative and negligible for all three fermions inside the loop for a large range of masses.

\subsection{Exotic Neutrino contributions}

\begin{figure}[H]
    \centering
    \includegraphics[scale=0.2]{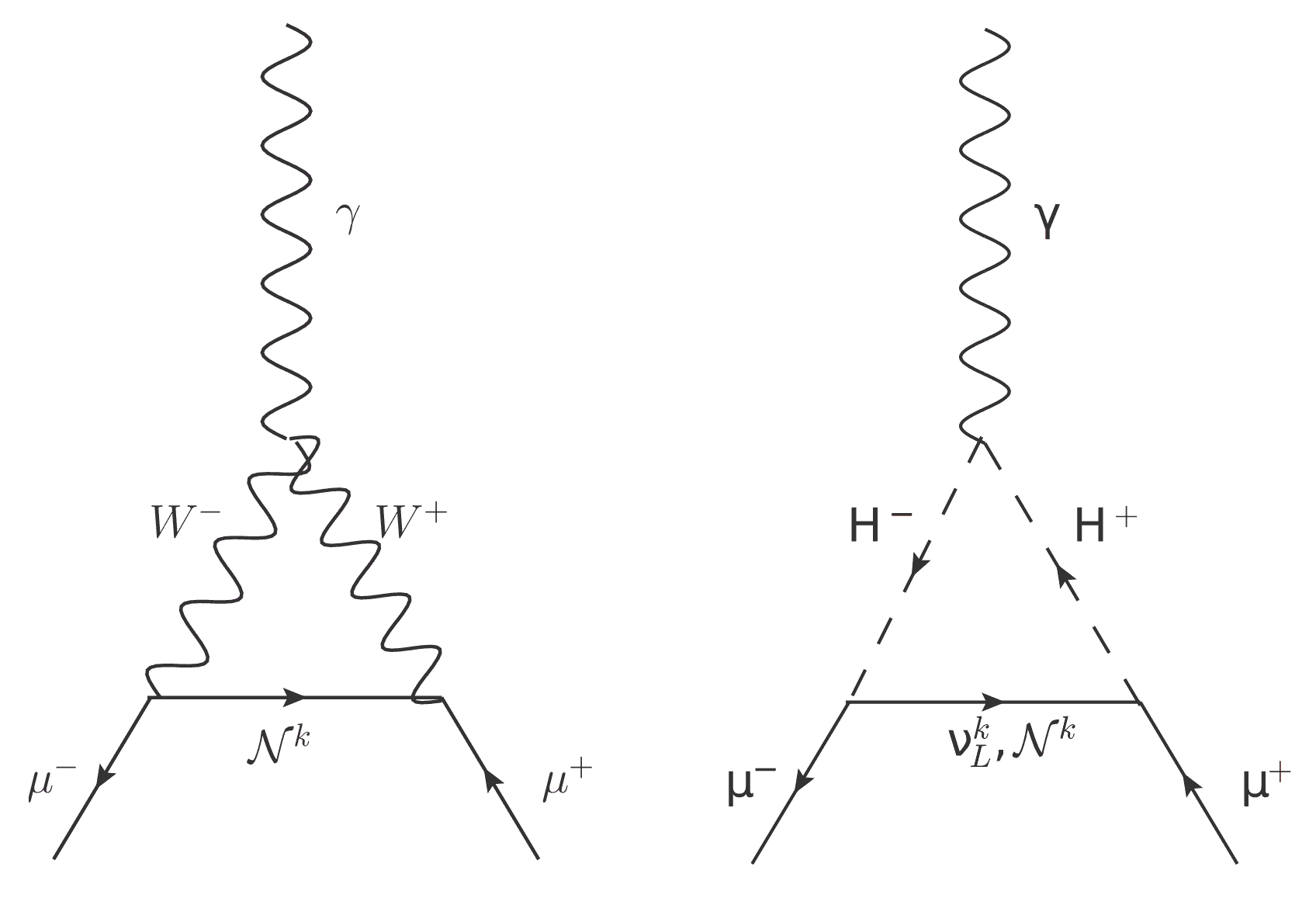}
    \caption{Contribution to muon $g-2$ due to exotic neutrinos into the loop interacting with charged (gauge)bosons.}
    \label{NuDiagrams}
\end{figure}
The interaction lagrangian for these processes is given by:
%\begin{align*}
%    \mathcal{L}_{int}&=-\bar{\nu}_{ex}^{j} W_{\mu} \frac{g_w  v_{2} \gamma^{\mu}P_{L}}{h_{N_{\chi k}}^3 v_{\chi}^3} \frac{\left(\mu_{N} \sqrt{2 h_{N_{\chi k}}^2 v_{\chi}^2+\mu_{N}^2}+h_{N_{\chi k}}^2 v_{\chi}^2+\mu_{N}^2\right)}{\sqrt{\left( \frac{\mu_{N}+\sqrt{\mu_{N}^2+2 h_{N_{\chi k}}^2 v_{\chi}^2}}{h_{N_{\chi k}} v_{\chi}}\right) ^2+2}}\left[ -s_{\theta_{e\mu}} h_{2e}^{\nu k} +c_{\theta_{e\mu}}h_{2\mu}^{\nu k} \right] \mu  + \nonumber \\
%    &+\bar{\nu}_{ex}^{j} H_{2}^{-}\Bigg[P_{R} \left(R_{3+j,1}^{\nu } (-s_{\beta} V_{2,4}^{R} h_{1e}^{E}+c_{\beta} V_{2,2}^{R} h_{2e}^{e\mu})+R_{3+j,2}^{\nu } (-s_{\beta} V_{2,4}^{R} h_{1\mu}^{E}+c_{\beta} V_{2,2}^{R} h_{2e}^{\mu\mu})\right)\\
%    &-c_{\beta} P_{L} \frac{\sqrt{2 h_{N_{\chi k}}^2 v_{\chi}^2+\mu_{N}^2}+\mu_{N}}{h_{N_{\chi k}} v_{\chi} \sqrt{\left( \frac{\mu_{N}+\sqrt{\mu_{N}^2+2 h_{N_{\chi k}}^2 v_{\chi}^2}}{h_{N_{\chi k}} v_{\chi}}\right)^2+2}}  \left(-s_{\theta_{e\mu}} h_{2e}^{\nu k }+c_{\theta_{e \mu}} h_{2\mu}^{\nu k }\right) \bigg]\mu + h.c.
%\end{align*}

\begin{align}
    \mathcal{L}_{int}&=-\bar{\mathcal{N}}^{j} W_{\mu}^{+} \frac{g_w  v_{2} \gamma^{\mu}P_{L}}{2h_{N_{\chi k}} v_{\chi}} \left[ -s_{\theta_{e\mu}} h_{2e}^{\nu k} +c_{\theta_{e\mu}}h_{2\mu}^{\nu k} \right] \mu  + \bar{\mathcal{N}}^{j} H_{2}^{+}\Bigg[ - P_{L} \frac{c_{\beta}}{\sqrt{2} }  \left(-s_{\theta_{e\mu}} h_{2e}^{\nu k }+c_{\theta_{e \mu}} h_{2\mu}^{\nu k }\right) \nonumber \\
    &+P_{R} \left(R_{3+j,1}^{\nu } (-s_{\beta} V_{2,4}^{R} h_{1e}^{E}+c_{\beta} V_{2,2}^{R} h_{2e}^{e\mu})+R_{3+j,2}^{\nu } (-s_{\beta} V_{2,4}^{R} h_{1\mu}^{E}+c_{\beta} V_{2,2}^{R} h_{2e}^{\mu\mu})\right)\bigg]\mu  + h.c. \label{Lnu}
\end{align}

where $k=1,2,1,2,3,3$ when $j=1,2,3,4,5,6$ respectively, which in the case of the Yukawa coupling it represents $k=e, \mu, e, \mu, \tau, \tau$. This is a consequence of the condition $\rho\sim 1$ which makes the degenerate lightest exotic neutrinos coming from the $h_{N_{\chi 1}}, h_{N_{\chi 2}}$ sectors and being the $h_{N_{\chi 3}}$ sector the heaviest. Besides, the interaction with charged scalars proportional to $P_{R}$ is not written explicitly because the contribution is too small ($g_{v}\sim 10^{-6}$) because the rotation matrix elements are related with seesaw rotations which are highly suppressed. However, the contribution due to the interaction with the $W^{\pm}$ boson is given by:

\begin{align}
\Delta a_{\mu} (W) &= \frac{1}{8\pi^2}\frac{m_\mu^2}{ M_{W}^2 } \int_0^1 dx \frac{|g_{v}|^2 \ P_{v}(x) + |g_{a}|^2 \ P_{a} (x) }{\epsilon^2 \lambda^2 (1-x)(1-\epsilon^{-2} x) + x},\\
&\approx \frac{1}{4\pi^2}\frac{m_\mu^2}{ M_{W}^2 } \left[|g_{v}|^2 \left( \frac{5}{6} - \frac{m_{\nu}}{m_{\mu}}\right)  + |g_{a}|^2 \left(  \frac{5}{6} + \frac{m_{\nu}}{m_{\mu}} \right) \right] \label{DeltaAnu}
\end{align}
where $\epsilon = m_{\nu}/m_{\mu}$ and $\lambda= m_{\mu}/M_{W}$, $g_{v}$ and $g_{a}$ the vector and axial couplings and 

\begin{eqnarray}
P_{v10}(x) & = &  2x^2(1+x-2\epsilon)+\lambda^2(1-\epsilon)^2 x(1-x)(x+\epsilon) \nonumber\\
P_{a10}(x) & = &  2x^2(1+x+2\epsilon)+\lambda^2(1+\epsilon)^2 x(1-x)(x-\epsilon),\nonumber\\
\end{eqnarray}

On the other hand, the contribution due to the interaction with charged scalars is given by:

\begin{align}
\Delta a_{\mu} (H^+) &= \frac{1}{8\pi^2}\frac{m_\mu^2}{ M_{H^+}^2 } \int_0^1 dx \frac{|g_{s}|^2 \ P_{s}(x) + |g_{p}|^2 \ P_{p} (x) }{\epsilon^2 \lambda^2 (1-x)(1-\epsilon^{-2} x) + x}\nonumber\\
&\approx \frac{1}{4\pi^2}\frac{m_\mu^2}{ M_{H^+}^2 }\left[  |g_{s}|^2\left(-\frac{ m_{\nu} }{4 m_{\mu}} -\frac{1}{12}\right) +  |g_{p}|^2 \left( \frac{ m_{\nu} }{4m_{\mu}} -\frac{1}{12}\right) \right]
\end{align}
where $\epsilon = m_{\nu}/m_{\mu}$, $\lambda= m_{\mu}/M_{H^+}$, $g_{s}$ and $g_{p}$ the scalar ans pseudo-scalar couplings respectively and
\begin{align}
P_{s2}(x) & =  -x (1-x)(x+\epsilon) & P_{p2}(x) & = -x (1-x)(x-\epsilon)
\end{align}

\begin{figure}[H]
     \centering
     \begin{subfigure}[b]{0.4\textwidth}
         \centering
         \includegraphics[scale=0.55]{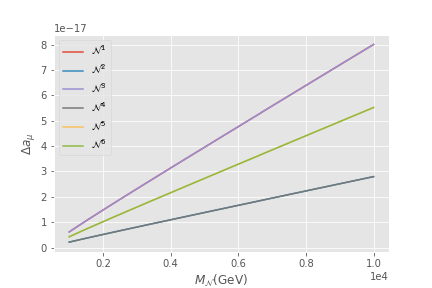}
         \caption{Contribution to $\Delta a$  due to exotic neutrinos interacting with $W^{\pm}$ gauge boson as a function of the exotic neutrino mass .}             \label{nucontributions-a}
     \end{subfigure}
     \hfill
     \begin{subfigure}[b]{0.5\textwidth}
         \centering
         \includegraphics[scale=0.55]{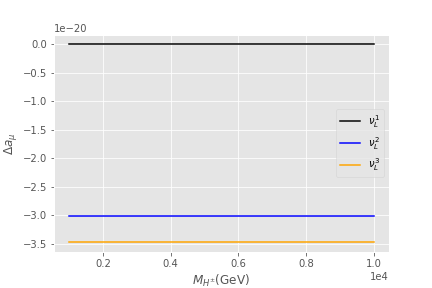}
         \caption{Contribution to $\Delta a$ from charged scalars and SM neutrinos as a function of the scalar mass.}
     \end{subfigure} \hfill
     \begin{subfigure}[b]{0.4\textwidth}
         \centering
         \includegraphics[scale=0.55]{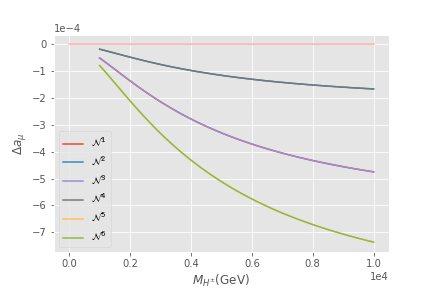}
         \caption{Contribution to $\Delta a$ due to exotic neutrinos itneracting with charged scalars as a function of the charged scalar mass for a neutrino of mass $1TeV$.}
     \end{subfigure}
     \hfill
     \begin{subfigure}[b]{0.5\textwidth}
         \centering
         \includegraphics[scale=0.55]{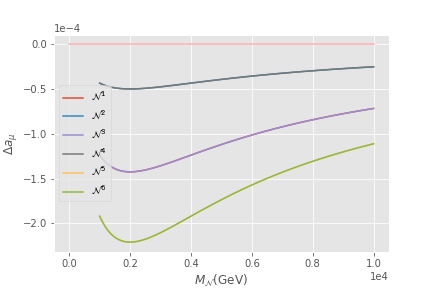}
         \caption{Contribution to $\Delta a$ due to exotic neutrinos interacting with charged scalars as a function of neutrino mass for a charged scalar mass of $2TeV$.}
     \end{subfigure}      
     \caption{Contributions to muon $g-2$ as a function of the charged scalar mass.}
     \label{nucontribution-1-NS}
\end{figure}

Since $v_{\chi}$ must have a very high value, all couplings with charged $W$ bosons are negligible, being some of them of order $\sim  10^{-7}$ and other of order $\sim 10^{-10}$ which make negligible contributions, as shown in figure \ref{nucontribution-1-NS}-a, and become even smaller when the interaction is related to charged scalars and SM neutrinos which leads to negligible contributions as well, mainly because of the small masses. All in all, there is no important contribution coming from these interactions. However, from exotic neutrinos interacting with charged scalars the coupling is of order $\sim  10^{-2}$ or higher, the contributions are shown in figures \ref{nucontribution-1-NS}-b and \ref{nucontribution-1-NS}-c.

Here the contributions of $\mathcal{N}^{1}$ and $\mathcal{N}^{3}$ overlap, just like $\mathcal{N}^{2}$, $\mathcal{N}^{4}$ and $\mathcal{N}^{5}$, $\mathcal{N}^{6}$.  Since exotic neutrinos make large negative contributions, it has to be counteracted with moderate charged scalar masses. Likewise, the contribution as a function of the neutrino mass is negative as well but it increases asymptotically to zero for large masses. They are shown in figure \ref{nucontribution-1-NS}-d.

\subsection{Neutral scalar particles}
In this case only heavy scalar interactions are taken into account because the lightest scalar is identified as the Higgs boson. Thus, its contributions is already included into the SM prediction. Their interaction lagrangian is given by:
%+  \sin\theta_{e\mu}\frac{t_{3}v_{2}^{4} }{2 m_{\mu}^{2}}\left(\Sigma_{11} h_{2 e}^{e\mu} h_{2e}^{\tau e}+\Sigma_{13}h_{2 e}^{e\mu} h_{2e}^{\tau\tau}\right)
\begin{align}
 % \mathcal{L}_{int}(H,H_{\chi}, A^{0}) &= \bar{\mu}H\frac{c_{12}}{\sqrt{2}} \left[-\sin\theta_{e\mu} h_{2e}^{e\mu}+\cos\theta_{e\mu} h_{2e}^{\mu\mu} \right]\mu -\bar{\mu}H_{\chi} \frac{\lambda _6 v_1}{4\sqrt{2} \lambda _3 v_{\chi }} \left[-\sin\theta_{e\mu} h_{2e}^{e\mu}+\cos\theta_{e\mu} h_{2e}^{\mu\mu} \right]\mu\nonumber \\
%  &-\bar{\mu}A^{0}\frac{c_{\beta}c_{\gamma}}{\sqrt{2}} \frac{v_{1}v_{2} \left(g_{\chi \mathcal{E}}^2 v_{\chi }^2+\mu_E^2\right) \left(h_{1e}^{E} h_{2 e}^{e\mu}+h_{1 \mu }^E h_{2 e}^{\mu \mu }\right)}{2 m_{E}^2 m_{\mathcal{E}}^2 } \left(-h_{1e}^{E}\sin\theta_{e\mu}+h_{1\mu}^{E} \cos\theta_{e\mu}\right)\mu \\
 \mathcal{L}_{int}(H,H_{\chi}, A^{0})  &\approx \bar{\mu}H\frac{c_{12}}{\sqrt{2}} \left[-\sin\theta_{e\mu} h_{2e}^{e\mu}+\cos\theta_{e\mu} h_{2e}^{\mu\mu} \right]\mu -\bar{\mu}H_{\chi} \frac{\lambda _6 v_1}{4\sqrt{2} \lambda _3 v_{\chi }} \left[-\sin\theta_{e\mu} h_{2e}^{e\mu}+\cos\theta_{e\mu} h_{2e}^{\mu\mu} \right]\mu\nonumber \\
  &+\bar{\mu}A^{0}i\gamma^{5}\frac{v_{1}v_{2} c_{\beta}c_{\gamma} \left(h_{1e}^{E} h_{2 e}^{e\mu}+h_{1 \mu }^E h_{2 e}^{\mu \mu }\right)}{\sqrt{2} m_{E}^2 } \left(-h_{1e}^{E}\sin\theta_{e\mu}+h_{1\mu}^{E} \cos\theta_{e\mu}\right)\mu
\end{align}

\begin{figure}[H]
     \centering
     \begin{subfigure}[b]{0.4\textwidth}
         \centering
         \includegraphics[scale=0.14]{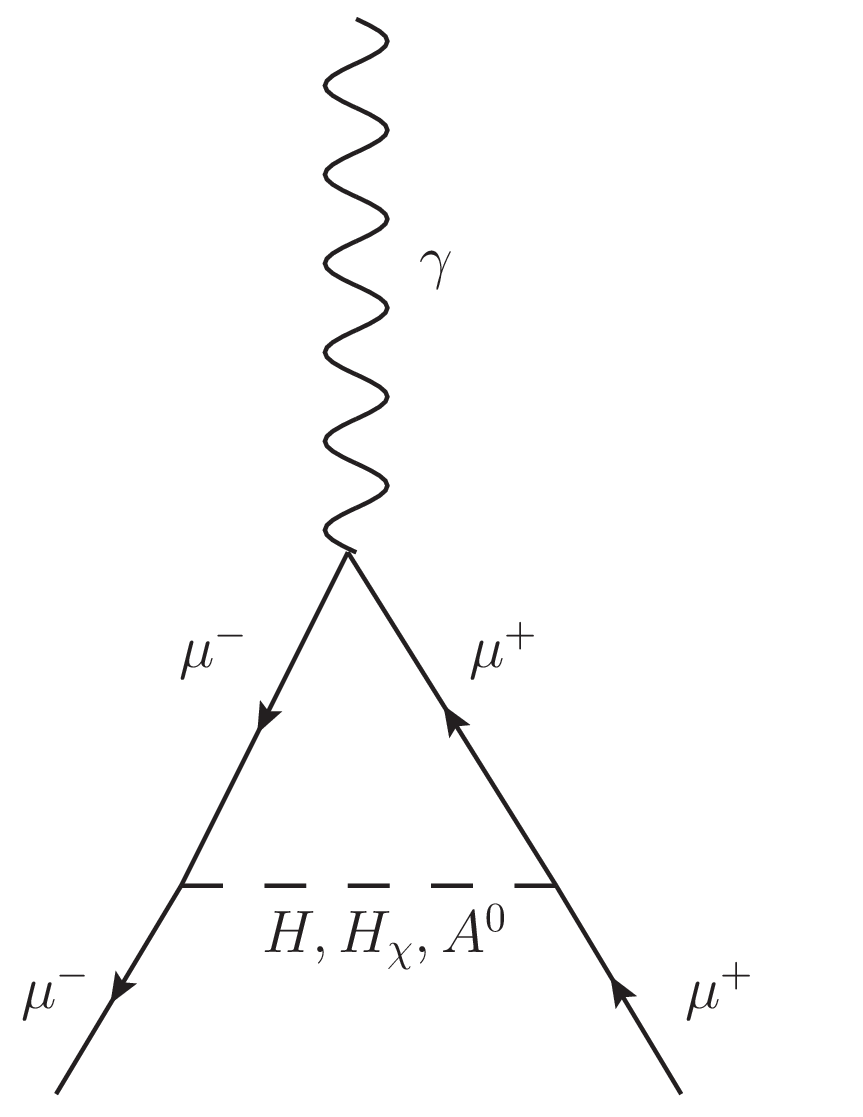}
         \caption{Feynman diagram for the (pseudo)scalar interaction with muons.}
     \end{subfigure}
     \hfill
     \begin{subfigure}[b]{0.5\textwidth}
         \centering
         \includegraphics[scale=0.6]{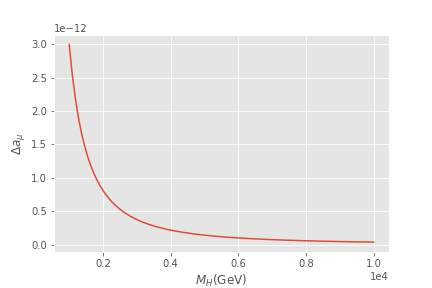}
         \caption{Contribution to $\Delta a$ from $H$ as a function of $H$ mass.}
     \end{subfigure}
     \hfill
     \begin{subfigure}[b]{0.5\textwidth}
         \centering
         \includegraphics[scale=0.6]{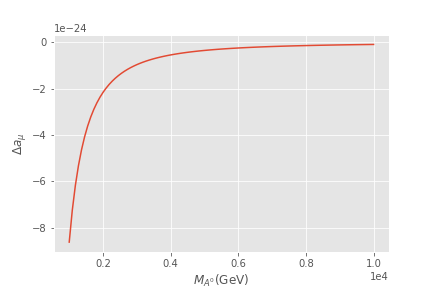}
         \caption{Contribution to $\Delta a$ from $A^{0}$ as a function of $A^{0}$ mass.}
     \end{subfigure}
     \caption{Scalar and pseudoscalar contributions to muon $g-2$.}
     \label{scalarcontributions}
\end{figure}

Scalar and Pseudoscalar particles provide separate contributions but they can be summarized as:

\begin{align}
\Delta a_{\mu} (\phi) &= \frac{1}{8\pi^2}\frac{m_\mu^2}{ M_{\phi}^2 } \int_0^1 dx \frac{g_{s}^2 \ P_{s}(x) + g_{p}^2 \ P_{p}(x) }{(1-x)(1-\lambda^2 x) +\lambda^2 x} \\
&\approx \frac{1}{4\pi^2}\frac{m_\mu^2}{ M_{\phi}^2 }\left[  g_{s}^2\left(  \ln \left(\frac{ M_{\phi} }{m_{\mu}} \right) -\frac{7}{12}\right) +  g_{p}^2\left( - \ln \left(\frac{ M_{\phi} }{m_{\mu}} \right) +\frac{11}{12}\right) \right]
\end{align}

where $\lambda=m_{\mu}/M_{\phi}$, $M_{\phi}$ represents the mass of the particle under consideration being $\phi=H,H_{\chi}, A^{0}$, $g_{s}$ and $g_{p}$ are the scalar and pseudo-scalar couplings and
\begin{align}
P_{s1}(x) & = x^2 (2 -x) & P_{p1}(x) & = - x^3,
\end{align}

since scalar and pseudoscalar particles are already distinguished in their mass eigenstate, scalar particles imply $g_{p}=0$ while pseudoscalar particles $g_{s}=0$. The contribution to muon $g-2$ due to $H$ and $A^{0}$ are shown in figure \ref{scalarcontributions} but contribution due to $H_{\chi}$ is not presented because its coupling is highly suppresed by $v_{\chi}$ as it can be seen in the interaction lagrangian, so it provides contributions smaller than $10^{-12}$.

We can see that the contribution due to both scalar and pseudoscalar are negligible. In the latter case, it is because of the null $(3,2)$ entry in the pseudoscalar rotation which suppress the contribution. Moreover, the contribution of the SM higgs boson is negligible but it is an expected result because its suppression is related to the Higgs to di-muon decay which is highly suppressed in the SM and represents less than the $0.02\%$ of the Higgs decays. Nonetheless, the present model agrees with the smallness of the observed signal since the interaction lagrangian for the model is given by:

%        $h_{4}$ & $-0.0000438348 $ & $\eta_{4}$ & $-0.0000438221$ \\ \hline
%        $h_{5}$ & $-0.00507714 $ &  $\eta_{5}$ & $-0.00507714 $ \\ \hline
%        $h_{6}$ & $0$ &  $\eta_{6}$ & $0$ \\ \hline
%        $h_{7}$ & $0$ & $\eta_{7}$ & $0$ \\ \hline
%        $h_{8}$ & $-6.70879 *10^{-11}$ &  $\eta_{8}$ & $6.70881*10^{-11}$ \\ \hline
%    \end{tabular}
%    \caption{Couplings with scalars and pseudoscalars. Their small values implies that there is no significant contribution.}
%    \label{tab:my_label}
%\end{table}

\begin{align}
    \mathcal{L}_{h\rightarrow \mu\bar{\mu}}&=\frac{v_{2}}{v}\frac{\sin\theta_{e\mu}h_{2e}^{2\mu}+\cos\theta_{e\mu}h_{2e}^{\mu\mu}}{\sqrt{2}} h\bar\mu\mu .
\end{align}

Consequently, the signal strength of the Higgs
decay rate into muon pair is modified from the SM prediction as:

\begin{align}
    \mu&=\frac{\Gamma(h\rightarrow \mu\mu)}{\Gamma(h\rightarrow \mu\mu)_{SM}} \\ \nonumber 
    &=\left|(\sin\theta_{e\mu}h_{2e}^{2\mu}+\cos\theta_{e\mu}h_{2e}^{\mu\mu} )\frac{v_{2}}{\sqrt{2}m_{\mu}}\right|^{2} \\\nonumber
    &=\sin^{2}(\theta_{e\mu}+\varphi)
\end{align}

being $\varphi$ an angle defined in parameter space by:
\begin{align}
    \tan\varphi &=\frac{h_{2e}^{\mu\mu}}{h_{2e}^{e\mu}}
\end{align}

The latest  experimental results from ATLAS \cite{htomumu} report a signal strength of $\mu=1.2 \pm 0.6$ corresponding to a $2\sigma$ significance in relation to the no-signal hypothesis. Since the model allows the parameter to be parametrized as the sine of an angle, it means that the model predicts a signal strength less or equal to the standard model prediction which is consistent with the current uncertainty. A graph of the allowed $\varphi$ and $\theta_{e\mu}$ angles in the $\sigma$ and $2\sigma$ intervals is shown:

\begin{figure}[H]
     \centering
     \begin{subfigure}[b]{0.4\textwidth}
         \centering
         \includegraphics[scale=0.6]{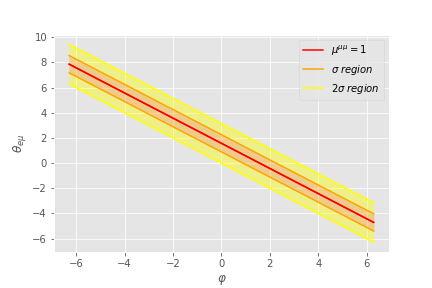}
         \caption{Allowed region for $\varphi$ and $\theta_{e\mu}$ consistent with the measured signal strength $\mu=1.2 \pm 0.6$ for a positive $\theta_{e\mu}-\varphi$ angle.}
     \end{subfigure}
     \hfill
     \begin{subfigure}[b]{0.5\textwidth}
         \centering
         \includegraphics[scale=0.6]{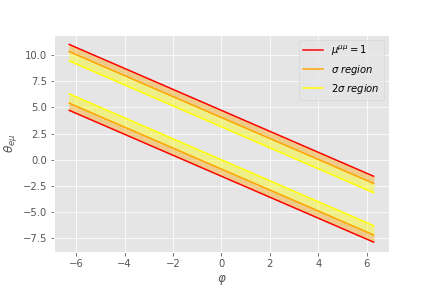}
         \caption{Allowed region for $\varphi$ and $\theta_{e\mu}$ consistent with the measured signal strength $\mu=1.2 \pm 0.6$ for a negative $\theta_{e\mu}-\varphi$ angle.}
     \end{subfigure}
\end{figure}

% the lower half region corresponds to the negative square root solution, and the $\phi$ interval is chosen to $(-2\pi,2\pi)$ to cover all posibilities

\section{The Supersymmetric extension} \label{SUSYgen}
We have seen that most of the contributions of the model are negligible, and those who are not are negative, related with the interaction of exotic neutrinos with charged scalars. Since there is no positive contribution that counteract it the model seems to be not consistent with the muon anomalous magnetic moment. Therefore, we explore the effect of imposing supersymmetry to the theory. When promoting fields to superfields the anomaly cancellation conditions are no longer true because now higgsinos enters have to be considered. Nevertheless, we can double the scalar superfield content in such a way that the new scalars behave as their conjugate and have the opposite quantum numbers as shown in table \ref{scalarlistSUSY}.
\begin{table}[H]
    \centering
    \begin{tabular}{|ccc|ccc|} \hline \hline
        Scalar Doublets & & & Scalar Singlets & &  \\
         & $X^{\pm}$ & $Y$ & & $X^{\pm}$ & $Y$   \\ \hline \hline
    $\small{\hat{\Phi}_{1}=\begin{pmatrix}\phi_{1}^{+}\\\frac{h_{1}+v_{1}+i\eta_{1}}{\sqrt{2}}\end{pmatrix}}$ & $\sfrac{+2}{3}^{+}$ & $+1$ & $\hat{\chi}=\frac{\xi_{\chi}+v_{\chi}+i\zeta_{\chi}}{\sqrt{2}}$	&	$\sfrac{-1}{3}^{+}$	&	$0$	\\
    $\small{\hat{\Phi}_{2}=\begin{pmatrix}\phi_{2}^{+}\\\frac{h_{2}+v_{2}+i\eta_{2}}{\sqrt{2}}\end{pmatrix}}$&$\sfrac{+1}{3}^{-}$&$+1$& $\hat{\sigma}=\frac{\sigma+i\zeta_{\sigma}}{\sqrt{2}} $ &$ \sfrac{-1}{3}^{-} $ & $ 0 $ \\\hline
    $\small{\hat{\Phi}'_{1}=\begin{pmatrix}\frac{h'_{1}+v'_{1}+i\eta'_{1}}{\sqrt{2}} \\ \phi_{1}^{\prime -}\end{pmatrix}}$ & $\sfrac{-2}{3}^{+}$ & $-1$ & $\hat{\chi}'=\frac{\xi'_{\chi}+v'_{\chi}+i\zeta'_{\chi}}{\sqrt{2}}$	&	$\sfrac{+1}{3}^{+}$	&	$0$	\\
    $\small{\hat{\Phi}'_{2}=\begin{pmatrix}\frac{h'_{2}+v'_{2}+i\eta'_{2}}{\sqrt{2}} \\ \phi_{2}^{\prime -}\end{pmatrix}}$&$\sfrac{-1}{3}^{-}$&$-1$& $\hat{\sigma}'=\frac{\sigma'+i\zeta'_{\sigma}}{\sqrt{2}} $ &$ \sfrac{+1}{3}^{-} $ & $ 0 $ \\\hline
    \end{tabular}
\caption{Model scalar particle content of the supersymmetric extension, $X$-charge, $\mathbb{Z}_{2}$ parity and hypercharge}
\label{scalarlistSUSY}
\end{table}

 Fermion fields can be promoted to superfields without any complication as right-handed superfields corresponds to conjugate fields $(f_{R}^{\dagger}\rightarrow \hat{f}_{L}^{c})$ which consequently have the opposite quantum number. In this way, the conjugate interaction lagrangian is considered when SUSY is imposed although the same mass matrix textures but charged leptons depends on primed VEVs while neutral leptons depend on non-primed VEVs. Moreover, gauge boson masses preserves its functional form, but the following replacements should be done:
\begin{align}
    v_{1} &\rightarrow \sqrt{v_{1}^{2} + v_{1}^{\prime 2}} &
    v_{2} &\rightarrow \sqrt{v_{2}^{2} + v_{2}^{\prime 2}}\\
    v_{\chi} &\rightarrow \sqrt{v_{\chi}^{2} + v_{\chi}^{\prime 2}} &
    v_{1}^{2}+v_{1}^{\prime 2}+v_{2}^{2}+v_{2}^{\prime 2} &=246^{2}
\end{align}

The only crucial difference lies on the scalar sector, because now we have 8 scalars, 8 pseudo-scalars and 4 charged scalars whose rotation matrices and interaction vertices are obtained numerically using FeynRules. Nonetheless, the interaction lagrangians are identical to the non-SUSY case except for the $c_{\beta}$, $s_{\beta}$ factors which have to be replaced by the corresponding matrix element. In this way, the interaction lagrangian for $Z'$ and $W^{\pm}$ interactions interactions are also true in the supersymmetric context after the VEV replacement is done. The following superpotential and soft breaking potential are considered:

\begin{align}
    W_{\phi}&=-\mu_{1}\hat{\Phi}'_{1}\hat{\Phi}_{1}-\mu_{2}\hat{\Phi}'_{2}\hat{\Phi}_{2} - \mu_{\chi}\hat{\chi} '\hat{\chi} - \mu_{\sigma}\hat{\sigma} '\hat{\sigma} + \lambda_{1}\hat{\Phi}_{1}^{\prime}\hat{\Phi}_{2}\hat{\sigma}^{\prime} + \lambda_{2}\hat{\Phi}_{2}^{\prime}\hat{\Phi}_{1}\sigma \label{4-9} \\[0.5cm]
    W_{L}&= \hat{\ell}_{L}^{p}\hat{\Phi_{2}}h_{2\nu}^{pq}\hat{\nu}_{L}^{q\; c} -\hat{\ell}_{L}^{p}\hat{\Phi '_{2}}{h}_{2e}^{p\mu}\hat{e}_{L}^{\mu\; c}
    - \hat{\ell}_{L}^{\tau}\hat{\Phi '_{2}}{h}_{2e}^{\tau r}\hat{e}_{L}^{r\; c} 
    - \hat{\ell}_{L}^{p}\hat{\Phi '_{1}}{h}_{1E}^{p}\hat{E}_{L}^{c}+ \hat{E}_{L}\hat{\chi '}{g}_{\chi' E}\hat{E}_{L}^{c}  \nonumber \\
    &- \hat{E}_{L}\mu_{E}\hat{\mathcal{E}}_{L}^{c} + \hat{\mathcal{E}}_{L}\hat{\chi}g_{\chi\mathcal{E}}\hat{\mathcal{E}}_{L}^{c}  - \hat{\mathcal{E}}_L\mu_{\mathcal{E}}\hat{E}_{L}^{c} +\hat{\nu}_{L}^{m\; c}\hat{\chi} ' {h}_{\chi}^{\prime N\; mn}\hat{N}_{L}^{n\; c}
        + \frac{1}{2}\hat{N}_{L}^{m\; c} M_{mn}\hat{N}_{L}^{n\; c} \nonumber\\
        &+ \hat{E}_{L}\hat{\sigma} h_{\sigma}^{e^{c} p}\hat{e}_{L}^{c r} + \hat{\mathcal{E}}_{L}\hat{\sigma}^{\prime}h_{\sigma'}^{e^{c} \mu}\hat{e}_{L}^{\mu c},\label{4-11}
\end{align}

where $j=1,2,3$ labels the down type singlet quarks, $k=1,3$ labels the first and third quark doublets, $a=1,2$ is the index of the exotic $\mathcal{J}_{L}^{a}$ and $\mathcal{J}_{L}^{c a}$ quarks, $p=e,\mu$ , $q=e,\mu,\tau$, $r=e,\tau$ and $m,n$ label the right handed and Majorana neutrinos. 

\begin{align}\label{Vs}
    V_{soft}&=m_{1}^{2}\Phi_{1}^{\dagger}\Phi_{1} + {m}_{1}^{\prime 2}{\Phi}_{ 1}^{\prime \dagger}\Phi'_{1} + m_{2}^{2}\Phi_{2}^{\dagger}\Phi_{2} + {m}_{2}^{\prime 2}\Phi _{2}^{\prime\dagger}\Phi'_{2}+m_{\chi}^{2}\chi^{\dagger}\chi + {m}_{\chi}^{\prime 2}{\chi}^{\prime\dagger}\chi'+m_{\sigma}^{2}\sigma^{\dagger}\sigma \nonumber\\
    & + {m}_{\sigma}^{\prime 2}{\sigma}^{\prime\dagger}\sigma'  -\bigg[\mu_{11}^{2}\epsilon_{ij}({\Phi}_{1}^{\prime i}\Phi_{1}^{j}) -\mu_{22}^{2}\epsilon_{ij}({\Phi}_{2}^{\prime i}\Phi_{2}^{j}) -\mu_{\chi\chi}^{2}(\chi\chi') +\mu_{\sigma\sigma}^{2}(\sigma\sigma') + \tilde{\lambda}_{1}\Phi_{1}^{\prime \dagger}\Phi_{2}\sigma^{\prime}\nonumber\\
    & + \tilde{\lambda}_{2}\Phi_{2}^{\prime \dagger}\Phi_{1}\sigma - \frac{2\sqrt{2}}{9}(k_{1}\Phi_{1}^{\dagger}\Phi_{2}\chi' -k_{2}\Phi_{1}^{\dagger}\Phi_{2}\chi^*+k_{3}\Phi_{1}'{}^{\dagger}\Phi_{2}'\chi -k_{4}\Phi_{1}'{}^{\dagger}\Phi_{2}'\chi'{}^*)+h.c.\bigg] \nonumber \\
    &+ M_{\tilde{B}}\tilde{B}\tilde{B}^{\dagger} + M_{\tilde{B}'}\tilde{B}'\tilde{B}^{\prime \dagger} + M_{\tilde{W}^{\pm}}\tilde{W}^{\pm}\tilde{B}^{\pm \dagger} + M_{\tilde{W}}\tilde{W}_{3}\tilde{W}_{3}^{\dagger} \\ \nonumber
\end{align}

As we saw in the last section, $v_{\chi}$ suppress the contributions to muon $g-2$ when we have neutrinos and $W$ bosons interacting into the loop. However, since we now have two $\chi$-VEVs and additional soft parameters the condition $v_{\chi}\sim 10^{7}$ can be relaxed to $v_{\chi}, v_{\chi}^{\prime } \sim 10^{3} \; GeV$.

\subsection{Scalar Sector}
Additional to the interactions coming from the superpotential, scalar particles receives contributions from $D-Terms$. Thus after SSB, the CP-even mass matrix, written in the basis $(h_{1},h'_{1},h_{2},h'_{2},\xi_{\chi},\xi'_{\chi},\xi_{\sigma},\xi'_{\sigma})$ is written as:

\begin{align}\label{h}
    \frac{1}{2}M_{h}^{2}&=\begin{pmatrix}
    M_{hh} & M_{h\xi}  \\
    M_{h\xi}^{T} & M_{\xi\xi} 
    \end{pmatrix},
\end{align}
where $M_{hh}$ is a $4\times 4$ matrix containing the mixing of the $h$ fields, related with the scalar doublets of the model. It can be written as:
\footnotesize
\begin{align}\label{mphi}
M&_{hh}=\nonumber\\
&\begin{pmatrix}
    f_{4g}v_{1}^{2}-\frac{v_{2}f_{1k}}{9v_{1}}+\frac{v_{1}'\mu_{11}^{2}}{2v_{1}} & -f_{4g}v_{1}v_{1}'-\frac{\mu_{11}^{2}}{2} &  f_{2g}v_{1}v_{2}+\frac{f_{1k}}{9}&-f_{2g}v_{1}v_{2}'+ \frac{1}{2}\lambda_{2}^{2}v_{1}v_{2}^{\prime}\\
    * & f_{4g}v_{1}'{}^{2}-\frac{v_{2}'f_{2k}}{9v_{1}'}+\frac{v_{1}\mu_{11}^{2}}{2v_{1}'}& -f_{2g}v_{1}'v_{2} + \frac{1}{2}\lambda_{1}^{2}v_{2}v_{1}^{\prime}&f_{2g}v_{1}'v_{2}'+\frac{f_{2k}}{9}\\
    *&*& f_{1g}v_{2}^{2}-\frac{v_{1}f_{1k}}{9v_{2}}+\frac{v_{2}'\mu_{22}^{2}}{2v_{2}}& -f_{1g}v_{2}v_{2}'-\frac{\mu_{22}^{2}}{2}\\
    *&*&*&f_{1g}v_{2}'{}^{2}-\frac{v_{1}'f_{2k}}{9v_{2}'}+\frac{v_{2}\mu_{22}^{2}}{2v_{2}'}
    \end{pmatrix}.
\end{align}

\normalsize
\noindent

It can be seen that the latter mixing matrix does not depend on the $m_{H\alpha}^{(\prime)}$ masses, they do not appear explicitly due to the minimum conditions stated above. As a consequence, the mixing is determined mainly by the $\mu_{ii}$ couplings, coming from the soft breaking potential rather than the superpotential parameters. However, the mixing between scalar doublets and singlets are written in the $4\times 4$ $M_{h\xi}$  matrix and it is given by:

\footnotesize	
\begin{align}
     M&_{h\xi}=\nonumber \\
     &\begin{pmatrix}
    \frac{1}{9}(k_{2}v_{2}-g_{X}^2v_{1}v_{\chi}) & \frac{1}{9}(-k_{1}v_{2}+g_{X}^2v_{1}v_{\chi}') & \frac{1}{2\sqrt{2}}(\tilde{\lambda_{2}}v_{2}^{\prime} - \lambda_{2}\mu_{2}v_{2}) & -\frac{1}{2\sqrt{2}}(\lambda_{1}\mu_{1}v_{2} + \lambda_{2}\mu_{\sigma}v_{2}^{\prime} ) \\
    \frac{1}{9}(-k_{3}v_{2}'+g_{X}^2v_{1}'v_{\chi}) & \frac{1}{9}(k_{4}v_{2}'-g_{X}^2v_{1}'v_{\chi}') &  -\frac{1}{2\sqrt{2}}(\lambda_{2}\mu_{1}v_{2}^{\prime} + \lambda_{1}\mu_{\sigma}v_{2}) & \frac{1}{2\sqrt{2}}(\tilde{\lambda_{1}}v_{2} - \lambda_{1}\mu_{2}v_{2}^{\prime})\\
    \frac{1}{9}(k_{2}v_{1}-\frac{1}{2}g_{X}^2v_{2}v_{\chi})& \frac{1}{9}(-k_{1}v_{1}+\frac{1}{2}g_{X}^2v_{2}v_{\chi}') &  -\frac{1}{2\sqrt{2}}(\lambda_{2}\mu_{2}v_{1} + \lambda_{1}\mu_{\sigma}v_{1}^{\prime}) & \frac{1}{2\sqrt{2}}(\tilde{\lambda_{1}}v_{1}^{\prime} - \lambda_{1}\mu_{1}v_{1})\\
    \frac{1}{9}(-k_{3}v_{1}'+\frac{1}{2}g_{X}^2v_{2}'v_{\chi})& \frac{1}{9}(k_{4}v_{1}'-\frac{1}{2}g_{X}^2v_{2}'v_{\chi}') & \frac{1}{2\sqrt{2}}(\tilde{\lambda_{2}}v_{1} - \lambda_{2}\mu_{1}v_{1}^{\prime}) & -\frac{1}{2\sqrt{2}}(\lambda_{1}\mu_{2}v_{1}^{\prime} + \lambda_{2}\mu_{\sigma}v_{1})
    \end{pmatrix}.
\end{align}

\normalsize
\noindent
We can see that the mixing between these two sectors, both expected at a different energy scale, is governed by the trilinear couplings of the soft breaking potential.  It implies that at the SUSY scale, scalar singlets and doublets are completely decoupled. Then, the SUSY breaking provides interactions between them and so the possibility of being observed at the right energy. Last but not least, the mixing matrix between Higgs singlets, $M_{\xi \xi}$, reads:

\footnotesize
\begin{align}\label{mxx}
    M_{\xi\xi}&=
    \begin{pmatrix}
   \frac{g_{X}^2}{18}v_{\chi}^2 +\frac{v_{\chi}'\mu_{\chi\chi}^2}{2v_{\chi}}-\frac{k_{23}}{9v_{\chi}} & -\frac{g_{X}^2}{18}v_{\chi}v_{\chi}'-\frac{\mu_{\chi\chi}^{2}}{2} & 0 & 0\\
    * &  \frac{g_{X}^2}{18}v_{\chi}'{}^2 +\frac{v_{\chi}\mu_{\chi\chi}^2}{2v_{\chi}'}-\frac{k_{14}}{9v_{\chi}'} & 0 & 0 \\
    * & * &M_{\sigma}^{2} +\frac{\lambda_{2}^{2}}{4}(v_{1}^{2}+v_{2}^{\prime 2})& -\frac{\mu_{\sigma\sigma}}{2} \\
    * & * & * &M_{\sigma}^{\prime 2} +\frac{\lambda_{1}^{2}}{4}(v_{2}^{2}+v_{1}^{\prime 2}) 
    \end{pmatrix} 
\end{align} 
\normalsize
\noindent

The following definitions have been done to give shorter expressions:
\begin{align}
    f_{ng}&=\frac{g^{2}+g'{}^{2}}{8}+\frac{n}{18}g_{X}^{2} & f_{1k}&=k_{2}v_{\chi}-k_{1}v_{\chi}' &  f_{2k}&=-k_{3}v_{\chi}+k_{4}v_{\chi}' \\
    k_{23}&=k_{2}v_{1}v_{2}-k_{3}v_{1}'v_{2}' & k_{14}&=-k_{1}v_{1}v_{2}+k_{4}v_{1}'v_{2}'
\end{align}
\begin{align}
    M_{\sigma}&=\frac{1}{2}(\mu_{\sigma}^{2} + m_{\sigma}^{2}) - \frac{g_{X}^{2}}{36}(2v_{1}^{2}+v_{2}^{2}-2v_{1}^{\prime 2}-v_{2}^{\prime 2} - v_{\chi}^{2} +v_{\chi}^{\prime 2}) \\ M_{\sigma}^{\prime}&=\frac{1}{2}(\mu_{\sigma}^{2} + m_{\sigma}^{\prime 2}) + \frac{g_{X}^{2}}{36}(2v_{1}^{2}+v_{2}^{2}-2v_{1}^{\prime 2}-v_{2}^{\prime 2} - v_{\chi}^{2} +v_{\chi}^{\prime 2})
\end{align}

The high energy decoupling of the doublet and singlet sectors lead us to assume the hierarchy $\mu_{\chi\chi}, \mu_{\sigma \sigma}, M_{\sigma}, M_{\sigma}^{\prime}\gg\mu_{11}\text{, }\mu_{22} \gg k_{i}v_{j}\gg g_{X}^{2}v_{\chi}v_{j}\text{, }g_{X}^{2}v'_{\chi}v_{j}\text{, }g_{X}^{2}v_{\chi}v'_{j}\text{, }g_{X}^{2}v'_{\chi}v'_{j}, \lambda_{i}^{2}v_{i}v_{j}^{(\prime)}$, where $i=1,2,3,4$ and $j=1,2$. Besides, no singlet has been observed so the $U(1)_{X}$ is expected at a much higher energy scale, implying that  $v_{\chi}$ and $v'_{\chi}$ should be at least at the TeV scale. Thus, they satisfy $v_{\chi}, v'_{\chi} \gg v_{j}, v'_{j}$, where $j=1,2$ . It implies for the mixing matrices $\mathcal{O}( M_{\xi\xi})\gg \mathcal{O}( M_{h\xi})\gg\mathcal{O}( M_{hh})$ which is a favorable scenario to implement a seesaw mechanism \cite{typeIseesaw}, with a rotation matrix $V$, leading to a block-diagonal form of the matrix represented by $\Tilde{M}_{h}^{2}$.

\begin{align}\label{hSS}
    \frac{1}{2}\Tilde{M}_{h}^{2}&=V\frac{1}{2}M_{h}^{2}V^{\dagger}\approx\begin{pmatrix}
    \Tilde{M}_{hh} & 0 \\
    0 & M_{\xi\xi}
    \end{pmatrix}, & V&=\begin{pmatrix}
    \mathbb{I} & M_{h\xi}M_{\xi\xi}^{-1} \\
    -(M_{h\xi}M_{\xi\xi}^{-1})^{T} & \mathbb{I} 
    \end{pmatrix}\nonumber
\end{align}

The matrix rank for the $M_{hh}$ submatrix is 4, which means that the four lightest eigenstates are massive and acquire their tree level mass from its mixing. Consequently, the seesaw contribution $M_{h\xi}M_{\xi\xi}^{-1}M_{h\xi}^{T}$ enters as small corrections to the tree level mass and can be neglected because of the order of magnitude of the involved parameters in each submmatrix. Thus, we can assume $\Tilde{M}_{hh} \approx M_{hh}$ and the block diagonal mass matrix takes the form:

\begin{align}
    \frac{1}{2}\Tilde{M}_{h}^{2}&\approx\begin{pmatrix}
    \Tilde{M}_{hh} & 0 \\
    0 & M_{\xi\xi}
    \end{pmatrix} 
\end{align}

In fact, all mass eigenstates are certainly massive since the mass matrix has rank 8 before and after the seesaw rotation as well as after the assumption of $\Tilde{M}_{hh}$. On the one hand, it is straightforward to get the scalar singlet masses since its $4\times 4$ submatrix has a block diagonal form because a mixing among $\chi$, $\chi'$ and $\sigma$ and $\sigma'$ is forbidden by gauge symmetry. In general, these masses depends on several free parameters and singlets VEVs so very large masses can be considered.

On the other hand, the eigenvalues coming from the $M_{hh}$ submatrix, two are expected to be function of the soft-SUSY breaking parameters $\mu_{11}$, $\mu_{22}$ while the others on  $k_{i}$ doublets VEVs because the latter must be identified with the SM Higgs particle. Heavy eigenstates are obtained by taking a small VEV approximation with the limit $v_{1},v_{2},v_{1}^{\prime},v_{2}^{\prime}\rightarrow 0$ on additive terms. It causes the matrix rank to decrease to 3, verifying the hypothesis of a electroweak dependent lightest eigenvalue. From this approximation the two heavy states arise from the reduced matrix:
\begin{align}\label{v=0}
M_{hh}(v_{i},v_{i}'\rightarrow 0)&=\begin{pmatrix}
\frac{\mu_{11}^2}{2}\frac{v_{1}^{\prime}}{v_{1}}    & -\frac{\mu_{11}^2}{2} &0&0\\
*                           &\frac{\mu_{11}^2}{2}\frac{v_{1}}{v_{1}^{\prime}}   &0&0\\
*                           &*          & \frac{\mu_{22}^2}{2}\frac{v_{2}^{\prime}}{v_{2}}&-\frac{\mu_{22}^2}{2}\\
*                           &*                      &*&\frac{\mu_{22}^2}{2}\frac{v_{2}}{v_{2}^{\prime}}
\end{pmatrix} 
\end{align}
giving as a result the tree level eigenvalues:

\begin{align}\label{h34}
    m_{h3}^{2}&\approx \mu_{11}^{2}\frac{v_{1}^{2}+v_{1}^{\prime 2}}{v_{1}v_{1}^{\prime}}, & m_{h4}^{2}&\approx \mu_{22}^{2}\frac{v_{2}^{2}+v_{2}^{\prime 2}}{v_{2}v_{2}^{\prime}}.
\end{align}
The next eigenvalue comes from approximating the exact solution of the matrix quartic order characteristic function, given by Ferrari's method \cite{Ferrari} in order to get a leading term for its mass. Considering only the terms proportional to $\mu_{11}^{2}\mu_{22}^{2}$ the eigenvalue becomes fully dependent on the $k_{i}$ parameters, and it reads:

\begin{align}\label{h2}
m_{h2}^{2}&\approx \frac{2v^{2}(v_{1}v_{2}(k_{1}v_{\chi}^{\prime}-k_{2}v_{\chi})+v_{1}^{\prime}v_{2}^{\prime}(k_{3}v_{\chi}-k_{4}v_{\chi}^{\prime}))}{9(v_{1}^{2}+v_{1}^{\prime 2})(v_{2}^{2}+v_{2}^{\prime 2})}.
\end{align}

Ferrari's method provide an equation for the lightest eigenvalue which is identified as the SM Higgs particle. However, the resulting expression becomes too complicated for doing approximations. Nevertheless, we consider the determinant dominant terms (proportional to $\mu_{11}\mu_{22}$) and divide by the previously obtained eigenvalues, then we get an expression for the SM Higgs boson as:

\begin{align}\label{Higgsobservedmass}
m_{h1}^{2}&\approx \frac{g_{X}^{2}(2v_{1}^{2}+v_{2}^{2}-2v_{1}^{\prime 2}-v_{2}^{\prime 2})^{2} }{9(v_{1}^{2}+v_{2}^{2}+v_{1}^{\prime 2}+v_{2}^{\prime 2})}+ \frac{ (g^{2}+g'{}^{2})(v_{1}^{2}+v_{2}^{2}-v_{1}^{\prime 2}-v_{2}^{\prime 2})^{2}}{4(v_{1}^{2}+v_{2}^{2}+v_{1}^{\prime 2}+v_{2}^{\prime 2})}
\end{align}
\normalsize
\noindent

Let's define the angles  $\tan^{2}\Tilde{\beta}=\frac{v_{1}^{2}+v_{2}^{2}}{v_{1}^{\prime2}+v_{2}^{\prime2}}$, $\tan\beta_{1}=\frac{v_{1}}{v'_{1}}$ and $\tan\beta_{2}=\frac{v_{2}}{v'_{2}}$ so Eq. (\ref{Higgsobservedmass}) equivalent to:

\begin{align}\label{shortHiggsmass}
    m_{h1}^{2}&=m_{Z}^{2}\left(cos^{2}2\Tilde{\beta}+\frac{4}{9}\frac{g_{X}^{2}}{g^{2}+g'{}^{2}}(cos2\beta_{1}+\cos 2\beta_{2})^{2}\right) \nonumber\\
    &\approx m_{Z}^{2}\cos^{2}2\Tilde{\beta} + \Delta m_{h}^{2}
\end{align}

The first thing to notice is that it depends only on the electroweak VEV's and the coupling constants as expected as well as there is no dependence on the new physics' energy scale implied by $v_\chi$ and $v'_{\chi}$ nor soft SUSY breaking parameters like  $\mu_{11}$ and $\mu_{22}$ which in general dominate the mass spectrum in SUSY theories. In fact, the theory with additional scalar singlets and D-terms due to supersymmetry, the correction term $\Delta m_{h}^{2}$ might be at the same tree level order but its experimental value is compatible with the NMSSM and USSM models. \\

In the case of CP-odd scalars and charged scalars, heavy eigenstates are obtained with the same procedure stated above, so at tree level, CP-even and CP-odd heavy masses are the same because of the $\mu_{11}$ and $\mu_{22}$ dominance while charged scalar masses are equal to the first four CP-even eigenvalues in a tree level approximation. This allows to relax the condition $v_{\chi} \sim 10^{7} \text{ GeV}$ because now the first heavy scalar rely on both singlet VEVs and $k_{i}$ parameters. In fact, it is found that:
\begin{align}
k_{i} &\sim 10^3 & 0 < \lambda_{i}, \tilde{\lambda}_{i} &< 10^{3}  \\
\mu_{11},\mu_{22} &> 10^{4} & \mu_{\chi\chi}, \mu_{\sigma \sigma}, M_{\sigma}, M_{\sigma}^{\prime} &> 10^{8} \\
v_{\chi},v'_{\chi} &> 10^{3},
\end{align}

\section{SUSY generalized model contributions } \label{SUSYcontributions}
Now that the supersymmetric extension has proven to be compatible with SM Higgs boson and lepton masses we execute a numerical exploration of the contribution to the muon anomalous magnetic moment. However, we recall that interaction lagrangians are  identical with the only difference of the scalar rotation matrices contributions. We have considered the following set of parameters that recreate the SM Higgs mass, lepton masses and the PMNS matrix.

\begin{align*}
  v_{1}&=195.68 \; GeV& v_{2}&=138.36 \; GeV& v_{1}^{\prime}&=52.577 \; GeV & v_{2}^{\prime}&=20.577 \; GeV \\
  v_{\chi}&=2900 \; GeV & v_{\chi}^{\prime}&=6600 \; GeV& g_{X}&=0.823 &   \Sigma_{11} &= 4.482\times 10^{-6} \\ 
  h_{2e}^{e\mu} &= 0.00556 &   \Sigma_{13} &= 0.00001363 &  h_{2e}^{\mu\mu} &= 0.00466 & h_{2e}^{\tau e} &= 0.09710 \\ 
  h_{2e}^{\tau\tau} &= 0.0740 &   h_{1e}^{E} &= 0.3245&   h_{1\mu}^{E} &= 0.1715 &     g_{\chi E} &= 0.8858& \\ 
  \mu_{E} &= 0.45177\; GeV &    \mu_{\mathcal{E}} &= 0.9758 \; GeV &   g_{\chi\mathcal{E}} &= 0.478 &   Re[h_{e\nu}^{ee}] &= 4.803 \\ 
 Re[h_{2e}^{\nu \mu}] &= 3.591 &  Re[h_{2e}^{\nu \tau}] &= 3.7743 &  Re[h_{2\mu}^{\nu e}] &= 4.722&  Re[h_{2\mu}^{\nu \mu}] &= 2.3248 \\  
 Re[h_{2\mu}^{\nu \tau}] &= -0.191 &  Im[h_{2e}^{\nu e}] &= -0.612 &  Im[h_{2e}^{\nu \mu}] &= 0.6554 &  Im[h_{2e}^{\nu \tau}] &= 0.0479 \\
 h_{N_{\chi 1}} &= 5 &  h_{N_{\chi 2}} &= 5 &  h_{N_{\chi 3}} &= 5.5 &   \mu_{N} &= 4.392\times10^{-10} \; GeV
\end{align*}

The chosen VEVs satisfy the condition $v_{1}^{2}+v_{2}^{2}+v_{1}^{\prime 2}+v_{2}^{\prime 2}=246^{2} \text{GeV}^{2}$ where $v_{1}$ has the greatest value because it couples to the top quark mass, while $v_{2}^{\prime}$ couples to bottom quarks and leptons.

\subsection{Exotic neutrino contribution}

\begin{figure}[H]
     \centering
     \begin{subfigure}[b]{0.4\textwidth}
         \centering
         \includegraphics[scale=0.55]{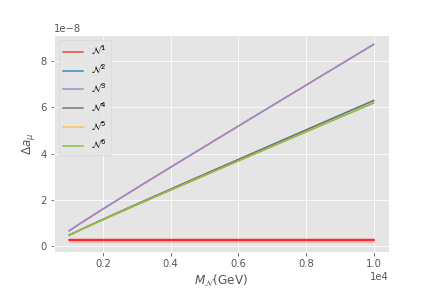}
         \caption{Contribution to $\Delta a$ due to exotic neutrinos interacting with $W^{\pm}$ gauge boson as a function of the exotic neutrino mass.}
        \label{nuwsusy}
     \end{subfigure}
     \hfill
     \begin{subfigure}[b]{0.4\textwidth}
         \centering
         \includegraphics[scale=0.55]{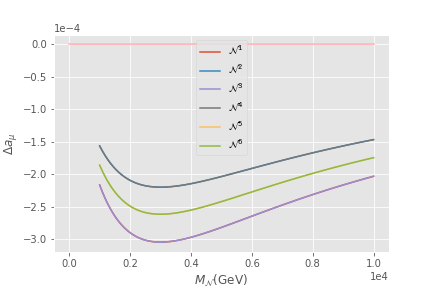}
         \caption{Contribution to $\Delta a$ due to exotic neutrinos interacting with charged scalars for a charged scalar of $3TeV$.}
     \end{subfigure}

     \caption{Contribution to muon $g-2$ due to exotic neutrinos for a charged scalar mass of $2$ TeV. The red line and its region represents the experimental value and its $1\sigma$ C.I. though it is small compared with the contributions.}
     \label{nucontribution-1}
\end{figure}

In this case, there is a relevant contribution from exotic neutrinos with masses above of $1TeV$ into the loop which has not been excluded yet \cite{exoticexclusion} as shown in figure \ref{nuwsusy}. In this case $g_{a}=-g_{v}$ and the contributions have a complex phase which is canceled because in the contribution to muon $g-2$ enters the coupling squared magnitude and the red line with their shadow corresponds to the latest measurement \cite{DeltaAmeasured} and its $1\sigma$ region. Because of the couplings, the $\mathcal{N}^{1}$ and $\mathcal{N}^{3}$, $\mathcal{N}^{2}$ and $\mathcal{N}^{4}$ and $\mathcal{N}^{5}$ and $\mathcal{N}^{6}$ curves overlap.

Since now $v_{\chi}$ can have a value of a few TeV, the contributions due to neutrinos interacting with $W$ gauge bosons is no longer negligible though they are considerably large. Their contributions are shown in figure \ref{nucontribution-1}-a while the contributions of exotic neutrinos interacting with charged scalars is shown in figure \ref{nucontribution-1}-b. Despite the interaction with charged scalars is still large and negative with the same behavior as in the non-SUSY case, we now have positive contributions in the first case that might explain the anomalous muon magnetic moment at larger masses.

%\begin{table}[H]
%    \centering
%    \begin{tabular}{|c|c|c|c|c|c|} \hline
%    Right neutrino & $g_{v}$& $\delta$ & Majorana Neutrino & $g_{v}$ & $\delta$ \\ \hline
%       $\nu_R^{1}$  & $-0.000461048e^{i\delta}$ & $-0.316679$ & $N_R^{1}$ & $-0.00343184e^{i\delta}$ & $0.041486$\\ \hline
%       $\nu_R^{2}$  & $-0.000461048e^{i\delta}$ & $0.316679$  & $N_R^{2}$ & $-0.00152474e^{i\delta}$ & $0.00491054$ \\ \hline
%       $\nu_R^{3}$  & $0.00343184e^{i\delta}$   & $0.041486$  & $N_R^{3}$ & $-0.00152474e^{i\delta}$& $0.00491054$ \\  \hline
%    \end{tabular}
%    \caption{Couplings with neutrinos, axial and vector couplings are related by $ig_{a}=g_{v}$.}
%\end{table}

%\begin{table}[H]
%    \centering
%    \begin{tabular}{|c|c|c|} \hline
%    Neutrino & $g_{v}$ & $\delta$ \\ \hline
%        $\nu_{L}^{1}$ & $1.62776*10^{-8}e^{i\delta}$& $0.0730949$ \\ \hline
%        $\nu_{L}^{2}$ & $-0.0000205074ie^{i\delta}$ & $0.318454$ \\ \hline
%        $\nu_{L}^{3}$ & $0.0000252531$ & $-0.0349542$ \\ \hline
%    \end{tabular}
%    \caption{Couplings with light neutrinos, axial and vector couplings are related by $-ig_{a}=g_{v}$.}
%\end{table}

%\begin{table}[H]
%    \centering
%    \begin{tabular}{|c|c|c|c|c|c|} \hline
%    Right neutrino & $g_{v}$& $\delta$ & Majorana Neutrino & $g_{v}$ & $\delta$ \\ \hline
%       $\nu_R^{1}$  & $-0.00152474e^{i\delta}$ & $0.00491054$ & $N_R^{1}$ & $1.45716e^{i\delta}$ & $0.041486$\\ \hline
%       $\nu_R^{2}$  & $0.195714e^{i\delta}$ & $0.316678$  & $N_R^{2}$ & $-0.712087e^{i\delta}$ & $0.00491054$ \\ \hline
%       $\nu_R^{3}$  & $1.45716e^{i\delta}$   & $0.041486$  & $N_R^{3}$ & $0.712087e^{i\delta}$& $0.00491054$ \\  \hline
%    \end{tabular}
%    \caption{Couplings with heavy neutrinos, axial and vector couplings are related by $ig_{a}=g_{v}$.}
%\end{table}

\subsection{SUSY contributions}

As expected from supersymmetry, there is a contribution due to charginos and neutralinos interacting with sneutrinos and sleptons respectively. There are in total 11 neutralinos and 3 charginos, however they rotation matrices take a rather complicated structure to diagonalize analytically so diagonalization was done numerically. The the interaction lagrangian can be approximated to: \\

\begin{align}
    \mathcal{L}_{int}&=-\bar{\tilde{\chi}}^{0}_{i} \Bigg[\frac{P_{R}}{3} \Big(\sqrt{2} (g_{X} R^{\tilde{\chi^{0}}}_{i,3} +  3 g' R^{\tilde{\chi^{0}}}_{i,1}) R^{\tilde{\ell}}_{7,j}+ 3 R^{\tilde{\chi^{0}}}_{i,7} (R^{\tilde{\ell}}_{j,1} h_{2e}^{e\mu} + R^{\tilde{\ell}}_{j,2} h_{2e}^{\mu\mu}) \Big) \\
   &-\frac{P_{L}}{6} \Big(6 R^{\tilde{\chi^{0}}}_{1,7} (-s_{\theta_{e\mu}} R^{\tilde{\ell}}_{7,1} h_{2e}^{e\mu} + c_{\theta_{e\mu}} R^{\tilde{\ell}}_{7,1} h_{2e}^{\mu\mu}) +      6 R^{\tilde{\chi^{0}}}_{1,5} R^{\tilde{\ell}}_{9,1} (-s_{\theta_{e\mu}} h_{1e}^{E} + c_{\theta_{e\mu}} h_{1\mu}^{E}) \nonumber \\
   &-3\sqrt{2}( g_{w} R^{\tilde{\chi^{0}}}_{1,2} + g' R^{\tilde{\chi^{0}}}_{1,1} ) (-s_{\theta_{e\mu}} R^{\tilde{\ell}}_{1,1} + c_{\theta_{e\mu}} R^{\tilde{\ell}}_{1,2}) \nonumber \\
   &- 2\sqrt{2} g_{X} R^{\tilde{\chi^{0}}}_{1,3} \Big(3  \frac{ s_{\theta_{e\mu}}v_{2}v'_{2}t_{3}}{2m_{\tau}^{2}}  R^{\tilde{\ell}}_{1,3} + 2 \frac{g_{\chi \mathcal{E} } v'_{1} v_{\chi}}{m_{E}m_{\mathcal{E}}}(h_{1e}^E s_{\theta_{e\mu}} -h_{1\mu}^E c_{\theta_{e\mu}}  ) ( 3 v_{\chi}R^{\tilde{\ell}}_{1,4} +2\mu_{\mathcal{E}}  R^{\tilde{\ell}}_{1,5}) \Big)  \Big) \Bigg] \tilde{\ell}_{j} \mu \nonumber\\
   & -\bar{\tilde{\chi}}^{+}_{i} \Bigg[P_{L}(g_{w} (-R^{\tilde{\nu}}_{j,1} s_{\theta_{e\mu}} + R^{\tilde{\nu}}_{j,2} c_{\theta_{e\mu}} ) R^{\tilde{\chi}^{+}}_{i,1} + ((R^{\tilde{\nu}}_{4,j} h_{2e}^{\nu e} + R^{\tilde{\nu}}_{5,j} h_{2e}^{\nu \mu} + R^{\tilde{\nu}}_{6,j} h_{2e}^{\nu\tau}) s_{\theta_{e\mu}} \nonumber \\
    &\;\;\;\;\;\;\;\;\;\;\;\;\;\;\;\;\;\;\;\;- (R^{\tilde{\nu}}_{4,j} h_{2\mu}^{\nu e} + R^{\tilde{\nu}}_{5,j} h_{2\mu}^{\nu\mu} + 
         R^{\tilde{\nu}}_{6,j} h_{2\mu}^{\nu\tau}) c_{\theta_{e\mu}}) R^{\tilde{\chi}^{+}}_{i,3}) +P_{R}R^{\tilde{\chi}^{-}}_{i,3} (R^{\tilde{\nu}}_{j,1} h_{2e}^{e\mu} + R^{\tilde{\nu}}_{j,2} h_{2e}^{\mu\mu}) \Bigg]\tilde{\nu}_{j} \mu
\end{align}

%missing term   3R^{\tilde{\chi^{0}}}_{1,11} R^{\tilde{\ell}}_{1,5} yex5x2

and the muon $g-2$ contributions are given by:
\begin{eqnarray}
&&
\Delta a_{\mu} (\phi) = \frac{1}{8\pi^2}\frac{m_\mu^2}{ M_{\tilde{\nu}}^2 } \int_0^1 dx \frac{g_{s1}^2 \ P_{s1}(x) + g_{p1}^2 \ P_{p1}(x) }{(1-x)(1-\lambda^2 x) +\lambda^2 x}
\end{eqnarray}where $\lambda=m_{\mu}/M_{\tilde{\nu}}$, $\epsilon=M_{\chi^{\pm}}/m_{\mu}$ and,
\begin{align}
P_{s1}(x) & = x^2 (2 +\epsilon-x) & P_{p1}(x) & = x^2 (2 -\epsilon-x),
\end{align}

\begin{figure}[H]
     \centering
     \begin{subfigure}[b]{0.4\textwidth}
         \centering
         \includegraphics[scale=0.15]{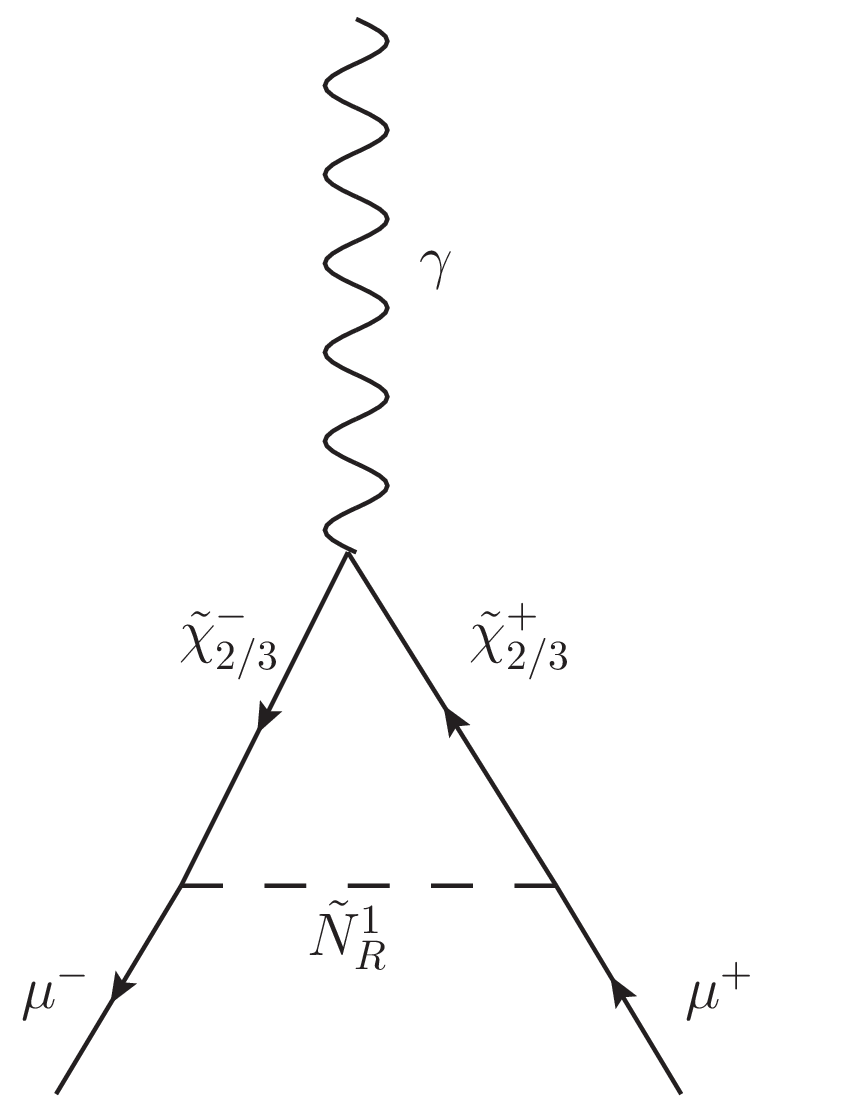}
         \caption{Feynman diagram representing for the contribution due to charginos and sneutrinos.}
     \end{subfigure}
     \hfill
     \begin{subfigure}[b]{0.5\textwidth}
         \centering
         \includegraphics[scale=0.6]{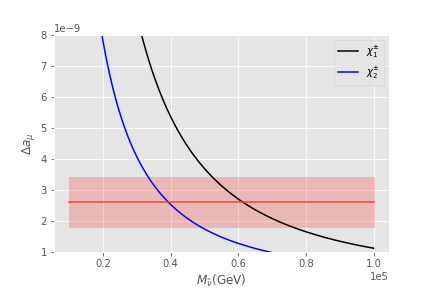}
         \caption{Contribution due to charginos and sneutrinos for a chargino mass of $1TeV$ as funtion of $\tilde{\nu}$ .}
     \end{subfigure} \hfill
     
     \begin{subfigure}[b]{0.4\textwidth}
         \centering
         \includegraphics[scale=0.6]{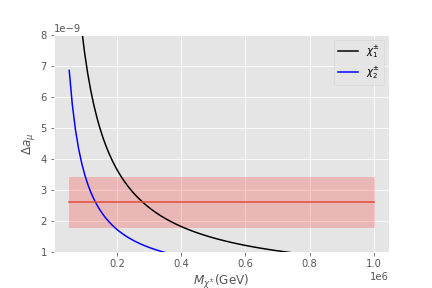}
         \caption{Contribution due to charginos and sneutrinos for a sneutrino mass of $3TeV$ as function of the chargino mass.}
     \end{subfigure}  \hfill
     \begin{subfigure}[b]{0.5\textwidth}
         \centering
         \includegraphics[scale=0.6]{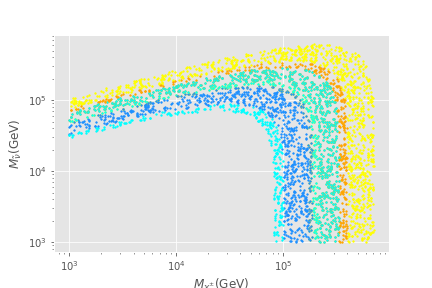}
         \caption{Mass region compatible with the experimental muon $g-2$ in a logarithmic scale. The orange and blue (yellow and cyan) region represent the $\sigma$ ($2\sigma$ confidence interval for the second and third chargino respectively. )}
     \end{subfigure}     
     \caption{Chargino-sneutrino contributions to muon $g-2$-}
     \label{charginos}
\end{figure}

In the case of charginos and sneutrinos the contribution to the muon $g-2$ is shown in figure \ref{charginos}. There are in total 27 non zero contributions but the couplings to the first chargino has small values, for that reason we focus our attention to a sample interaction term between second and third chargino which shows the general behavior of the contributions. It is presented as a function of the sneutrino mass in figure \ref{charginos}-b and as a function of chargino masses in figure \ref{charginos}-c.

We can see that the SUSY contributions have a decreasing character with mass, reaching the experimental value for masses of order $\sim 10^{5}\; GeV$. Additionally, a simple Monte-Carlo exploration shows the mass region compatible with the muon $g-2$ experimental value in figure \ref{charginos}-d for the same sample particles showing its compatibility with large sneutrino and chargino masses.

%\begin{table}[H]
%    \centering
%    \begin{tabular}{|c|c|c|} \hline
%        Particle & $g_{s}$ & $\delta$\\ \hline
%        $\Tilde{\chi}_{2}$ & $2.05734e^{i\delta}$ & $0.00624554$\\ \hline
%        $\Tilde{\chi}_{3}$ & $1.41389e^{i\delta}$ & $0.00624554$ \\ \hline
%    \end{tabular}
 %   \caption{Vector couplings for chargino-sneutrino interactions where $-ig_{p}=g_{s}$}
%\end{table}

\begin{figure}[H]
     \centering
     \begin{subfigure}[b]{0.4\textwidth}
         \centering
         \includegraphics[scale=0.55]{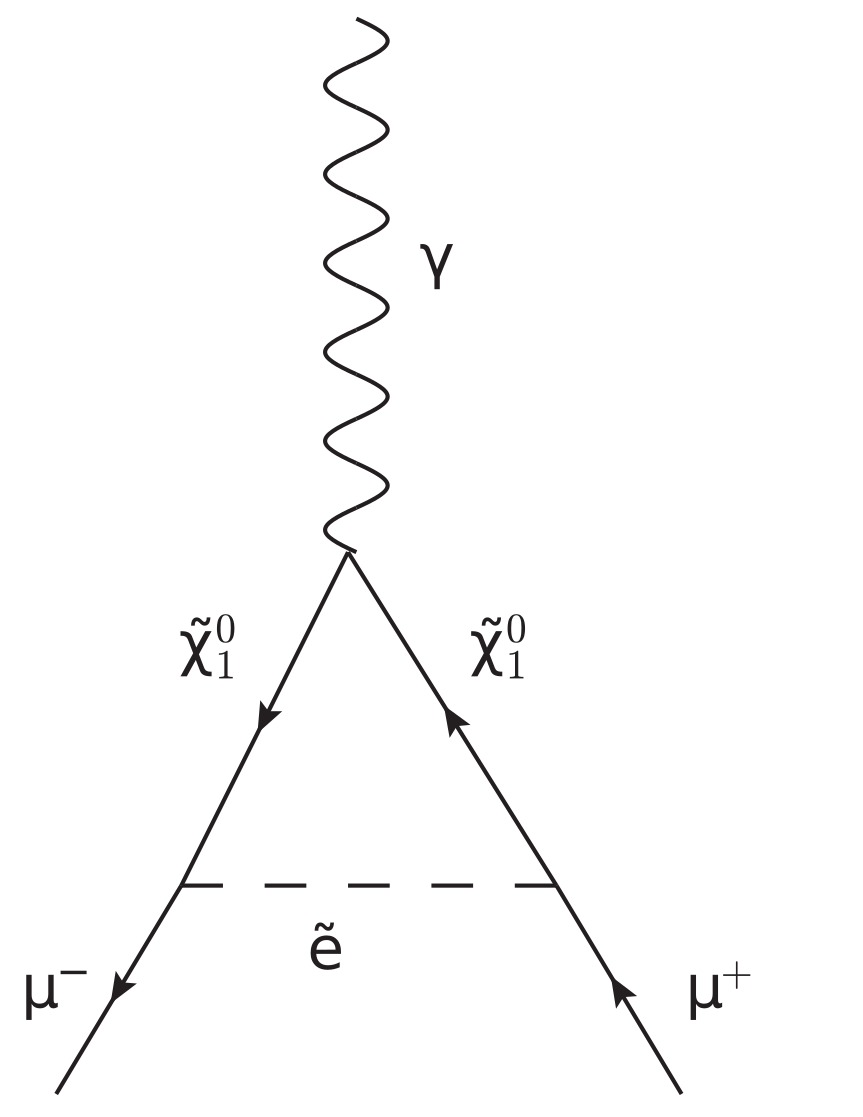}
         \caption{Feynman diagram for the contribution to muon $g-2$ due to neutralinos and sleptons into the loop.}
     \end{subfigure}
     \hfill
     \begin{subfigure}[b]{0.5\textwidth}
         \centering
         \includegraphics[scale=0.6]{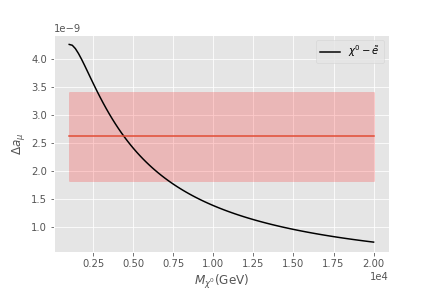}
         \caption{Contribution due to the lightest neutralino and selectron as a function of the neutralino mass.}
     \end{subfigure}      \hfill
     \begin{subfigure}[b]{0.5\textwidth}
         \centering
         \includegraphics[scale=0.6]{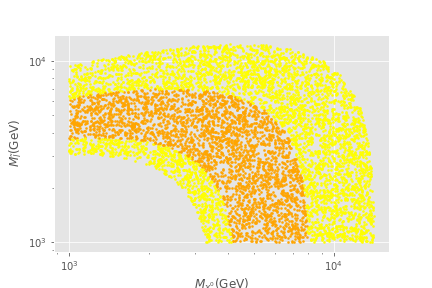}
         \caption{Mass region compatible with the experimental muon $g-2$ in a logarithmic scale. The orange(yellow) region represent the $\sigma$ ($2\sigma$ confidence interval).}
     \end{subfigure} 
     \caption{$\tilde{\chi}^{0}-\tilde{e}$ contribution to muon $g-2$.}
     \label{neutralinos}
\end{figure}

In the case of neutralinos, 90 out of 100 contributions are non zero with similar couplings of order $\sim 10^{-2}$. We consider the the case of the lightest neutralino $\chi_{1}^{0}$ with selectron $\tilde{e}$ in figure \ref{neutralinos}-b which shows a similar behavior of chargino-sneutrino interaction but now the contribution reach the experimental value for masses of order $\sim 10^{4}$ GeV.

%where the couplings are given by:
%\begin{align}
%    g_{s}&=0.104878 & g_{p}&=-0.104877i
%\end{align}

Due to the big amount of contributions we can consider that charginos and neutralinos have masses big enough to make their contributions negligible.so one possibility is neutralinos to be must much heavier than expected making negligible all contributions, yet it is presented in figure \ref{neutralinos}-c the mass region compatible with the observed $\Delta a_{\mu}$. 

\subsection{Cancellation of negative contributions}
\begin{figure}[H]
     \centering
     \begin{subfigure}[b]{0.4\textwidth}
         \centering
         \includegraphics[scale=0.6]{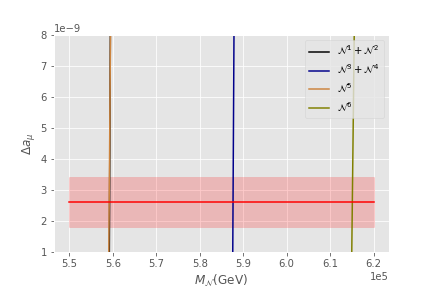}
         \caption{Added contributions of exotic neutrinos with chargino interactions for denegerate chargino masses, a charged scalar of $2$TeV and a sneutrino of $3$TeV mass.}
     \end{subfigure}
     \hfill
     \begin{subfigure}[b]{0.5\textwidth}
         \centering
         \includegraphics[scale=0.6]{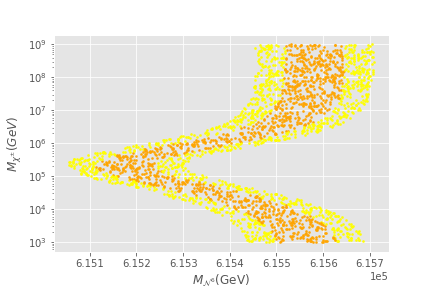}
         \caption{Parameter region of chargino and exotic neutrino $N^{3}$ mass compatible with the experimental muon $g-2$ value at $1\sigma$(orange) and $2\sigma$ (yellow) C.L.}
     \end{subfigure}
     \caption{Final contribution to muon $g-2$ in the supersymmetric theory and the allowed mass parameter region for degenerate chargino masses.}
     \label{Cancellation}
     \end{figure}

To show that added contributions can explain the experimental muon $g-2$ deviation, we add both exotic neutrino contributions to chargino-sneutrino ones by assuming neutrino and charginos of equal masses as shown in figure \ref{Cancellation}-a, for a sneutrino mass of $3$ TeV and a charged scalar of $2$TeV. In fact, for the order of magnitude where the cancellation is successful chargino contributions are negligible as might be concluded from figure \ref{charginos}-b and \ref{charginos}-c and in a similar fashion for neutralino contributions from figure \ref{neutralinos}-b. Besides, the graph does not show any significant difference when chargino contributions are excluded. Furthermore, in figure \ref{Cancellation}-b is shown the chargino and exotic neutrino masses compatible with experiments for the heaviest exotic neutrino. 

Only the $N^{3}$ parameter region is shown because the other two contributions have a similar graph and it shows that the experimental muon $g-2$ value can be explained in the context of a supersymmetric theory. Since each individual contribution shown in figure \ref{Cancellation}-a might explain the anomaly, when summing up all contributions is just a matter of doing the right choice for each exotic neutrino. This result shows that when all contributions are considered, exotic neutrinos are enough to  provide an explanation  while $v_{\chi}$ lies on the TeV scale.

\section{Conclusions}\label{conclu}
We have built an anomaly free model which is able to recreate the Standard Model Higgs boson mass with additional scalar particles lying at the TeV scale if $v_{\chi}\sim 10^{7}$ GeV. Additionally, SM lepton masses are recreated with the condition that the electron must acquire a finite mass at one loop leve, thanks to the interaction with the inert scalar $\sigma$ because the $U(1)_{X}\times \mathcal{Z}_{2}$ symmetry causes a tree-level massless electron. Likewise, the lightest neutrino is tree-level massless but no corrections are considered since we are interested in mass differences which make the latter negligible. Besides, a massless neutrino implies that neutrino mass spectrum can be fully determined prior to make a parameter fitting compatible with the PMNS matrix. In relation to muon $g-2$, it generates relevant negative contributions through exotic neutrinos in their mass basis interacting with heavy scalar particles into the loop while positive contributions arise when neutrinos interact with the SM $W$ boson is negligible due to the elevated value of $v_{\chi}$. Thus, the supersymmetric scenario was explored where again a SM like scalar particle was found and parity-violating interaction that breaks softly supersymmetry allows to consider $v_{\chi} \sim 10^{3}$ GeV causing considerable contributions due to the neutrino-$W^{\pm}$ interaction. All in all, a parameter region compatible with the experimental $\Delta a_{\mu}$ deviation is shown which demonstrates that positive and negative contributions cancel to explain the anomaly for masses of order $10^{5}$ GeV while contributions due to charginos and neutralinos are large in number but should be negligible according to higher masses.

\appendix
\section{Inverse seesaw block diagonalization}\label{appendinx}
Inverse seesaw rotation arise for a particular mass matrix texture given by:
\begin{align}
    \mathcal{M}&=\begin{pmatrix}
        0 & m^{T} & 0 \\
        m & 0 & M^{T} \\
        0 & M & \mu
    \end{pmatrix},
\end{align}
where in general $m$, $M$ and $\mu$ are not square matrices but their entries order or magnitude imply the hierarchy $\mu \ll m \ll M$. That matrix is symmetric but not generally hermitian diagonalization is done by the singular value decomposition. The inverse seesaw rotation goal is to rotate the latter matrix into a block diagonal form so each one of then can be independently diagonalized. First, the following blocks are defined: 

\begin{align}
    m_{D}&=\begin{pmatrix}
        m \\
        0
    \end{pmatrix}, & M_{heavy}&=\begin{pmatrix}
        0 & M^{T} \\
        M & \mu
    \end{pmatrix},
\end{align}
so  the original matrix can be rewritten as:
\begin{align}
    \mathcal{M}&=\begin{pmatrix}
        0 & m_{D}^{T} \\
        m_{D} & M_{heavy}
    \end{pmatrix}
\end{align}.
The latter matrix can be block diagonalized by an unitary matrix given by:
\begin{align}
    \mathcal{R}_{SS}&=\begin{pmatrix}
        \mathbb{I} & F^{T} \\
        -F & \mathbb{I}
    \end{pmatrix},
\end{align}
where the matrix basis denoted by $\mathbb{N}$ transform as $\mathcal{R}^{\dagger}\mathbb{N}$. Thus, the rotated matrix can be written as:
\begin{align}
    \mathcal{M}_{diag}&=\mathcal{R}^{T}_{SS}\mathcal{M}\mathcal{R}_{SS} \\
    &=\begin{pmatrix}
        -m_{D}^{T} F - F^{T}m_{D} + F^{T}M_{heavy} F & -F^{T}M_{heavy} F + m_{D}^{T} - F^{T}M_{heavy} \\
        -F m_{D}^{T} F + m_{D} - M_{heavy} F & Fm_{D}^{T} + m_{D}F^{T} + M_{heavy}
    \end{pmatrix} \\
    &\equiv \begin{pmatrix}
        m_{light} & 0 \\
        0 & \mathcal{M}_{H}
    \end{pmatrix}.
\end{align}

Thus, if we consider that $F$ is small we can neglect quadratic terms on it so an expression can be obtained from the off-diagonal elements as:
\begin{align}
    -F^{T}M_{heavy} F + m_{D}^{T} - F^{T}M_{heavy} &\approx m_{D}^{T} - F^{T}M_{heavy} \\
    F&=M_{heavy}^{-1}m_{D}
\end{align}
so the rotated matrix becomes:

\begin{align}
    \mathcal{M}_{diag}&\approx \begin{pmatrix}
        -m_{D}^{T} M_{heavy}^{-1}m_{D} & 0 \\
         0 & M_{heavy}^{-1}m_{D} m_{D}^{T} + m_{D}m_{D}^{T}M_{heavy}^{-1} + M_{heavy}        
    \end{pmatrix},
\end{align}
where

\begin{align}
    M_{heavy}^{-1}&=\begin{pmatrix}
        -M^{-1}\mu (M^{T})^{-1} & M^{-1} \\
        (M^{T})^{-1} & 0
    \end{pmatrix}.
\end{align}
The matrix $M_{heavy}^{-1}$ can be easily obtained and due to the submatrices hierarchy the terms proportional to $M_{heavy}^{-1}m_{D}$ are negligible small. Then, the block diagonal matrix and the F matrix rotation parameter are given by:
\begin{align}
    \mathcal{M}_{diag}&=\begin{pmatrix}
       m^{T}M^{-1}\mu (M^{T})^{-1}m & 0 \\
        0 & M_{heavy}
    \end{pmatrix}, & F&=\begin{pmatrix}
        -M^{-1}\mu (M^{T})^{-1} m \\
        (M^{T})^{-1}m \\
    \end{pmatrix},
\end{align}
so finally the above matrices are decoupled and can be diagonalized independently. The submatrix $m_{light}=m^{T}M^{-1}\mu (M^{T})^{-1}m$ contains the light states and is diagonalized by a rotation
\begin{align}
    \mathcal{R}_{\nu}&=\begin{pmatrix}
        V_{\nu} & 0 \\
        0 & \mathbb{I}
    \end{pmatrix}.
\end{align}
Nevertheless, Heavy states can be diagonalized by $V_{N}^{T}M_{heavy}V_{N}$ being $V_{N}$ a $\sfrac{\pi}{4}$ rotation followed by a seesaw rotation with parameters $S$, the result of the rotations should provide a block diagonal form as follows:
\begin{align}
    V_{N}&=\frac{1}{\sqrt{2}}\begin{pmatrix}
        \mathbb{I} & -\mathbb{I} \\
        \mathbb{I} & \mathbb{I} 
    \end{pmatrix}\begin{pmatrix}
        \mathbb{I} & S \\
        -S^{T} & \mathbb{I}
    \end{pmatrix}=\frac{1}{\sqrt{2}}\begin{pmatrix}
        \mathbb{I}-S^{T} & \mathbb{I} + S \\
        -(\mathbb{I} + S^{T}) & \mathbb{I} - S 
    \end{pmatrix}
    \approx
    \frac{1}{\sqrt{2}}\begin{pmatrix}
        \mathbb{I} & \mathbb{I} \\
        -\mathbb{I} & \mathbb{I}
    \end{pmatrix}
\end{align}
\begin{align}
    V_{N}^{T}M_{heavy}V_{N}&=\frac{1}{2} 
     \left(  \begin{array}{cc}
        \mathbb{I}  & -S \\
         S^T& \mathbb{I}
     \end{array}   \right)\left( \begin{array}{c|c}\small
            -M^T-M+\mu& M^T-M-\mu \\\hline
            -M^T+ M-\mu& M^T+ M+ M_{M} 
        \end{array}  \right)
     \left(  \begin{array}{cc}\small
        \mathbb{I}  & S \\
         -S^T & \mathbb{I}
     \end{array}   \right) \\
     &\equiv\begin{pmatrix}
         M_{1} &0 \\
         0 & M_{2}
     \end{pmatrix}.
\end{align}
By doing the matrix product and considering $M$, $\mu$, $M S^{\dagger}$ and $\mu S^{\dagger}$ to be symmetric matrices, the parameter $S$ can be obtained from the off-diagonal entries as:
\begin{align}
    S&=-\frac{1}{4}M^{-1}\mu
\end{align}
so after neglecting terms according to the parameter hierarchy and quadratic terms in S, the rotated matrix becomes:
\begin{align}
    V_{N}^{T}M_{heavy}V_{N}&\approx\begin{pmatrix}
        -M + \frac{1}{2}\mu & 0 \\
        0 & M + \frac{1}{2}\mu
    \end{pmatrix},
\end{align}
 with the corresponding rotation written as:
\begin{align}
    \mathcal{R}_{N}&=\begin{pmatrix}
        \mathbb{I} & 0 \\
        0 & V_{N}
    \end{pmatrix}.
\end{align}
In general there are two additional rotations taht should diagonalize $M_{1}$ and $M_{2}$. However, for simplicity we can be taken as diagonal matrices diagonalization is complete. All in all, the total rotation matrix that diagonalizes the original matrix $\mathcal{M}$ can be written as:

\begin{align}
\mathcal{R}&=\mathcal{R}_{SS}\mathcal{R}_{\nu}\mathcal{R}_{N} \\
&\approx\left( \begin{array}{c|cc}
         V_{\nu}& m^TM^{-1} &m^TM^{-1}\\\hline
        - M^{-1} \mu (M^T)^{-1} m V_\nu &\frac{1}{\sqrt{2}}\mathbb{I} & \frac{1}{\sqrt{2}}\mathbb{I}\\
         (M^T)^{-1}mV_{\nu}&-\frac{1}{\sqrt{2}}\mathbb{I} &\frac{1}{\sqrt{2}}\mathbb{I}
    \end{array} \right).
\end{align}

\end{document}